\documentclass[acmtog]{acmart} 
\acmSubmissionID{0270}

\usepackage{booktabs} % For formal tables
\setcitestyle{square}
\usepackage{ifthen}
\usepackage{enumitem}
\usepackage{algorithm}
\usepackage[noend]{algpseudocode}
\usepackage{syntax}
\usepackage{amsfonts}
\usepackage{listings}
\usepackage{fancyvrb}
\usepackage{wrapfig}
\usepackage{graphicx}
\usepackage{subfigure}
\usepackage{xspace}
\usepackage{colortbl}
\usepackage{gensymb}
\usepackage{verbatim}
\usepackage{colortbl}
\usepackage{booktabs}
\usepackage{multirow}

\algdef{SE}[DOWHILE]{Do}{doWhile}{\algorithmicdo}[1]{\algorithmicwhile\ #1}%
\definecolor{mygreen}{rgb}{0,0.6,0}                         
\definecolor{mygray}{rgb}{0.95,0.95,0.95}
\definecolor{codebg}{rgb}{0.95, 0.95, 0.95}  
\newcommand{\etal}{et al.}

\newcommand{\hs}[1]{{\bfseries \color{red} HZ: #1}}

\usepackage[normalem]{ulem}
\newcommand{\revision}[1]{\textcolor{black}{#1}}

\renewcommand{\grammarlabel}[2]{#1\hfill#2} 
\definecolor{mypink}{rgb}{.99,.91,.95}

\begin{document}
%\title{Extrusion-Based Ceramics Printing of Shell Models}
%\title{As-Continuous-As-Possible Ceramics Printing for Shell Models}
\title{\revision{As-Continuous-As-Possible Extrusion Fabrication of Surface Models}}

\author{Fanchao Zhong}
\email{fanchao98@gmail.com}
\affiliation{%
  \institution{Shandong University}
}

\author{Yonglai Xu}
\email{xyliyrwi@gmail.com}
\affiliation{%
  \institution{Shandong University}
}

\author{Haisen Zhao}
\email{haisen.zhao@ist.ac.at}
\affiliation{%
  \institution{IST Austria, Shandong University and University of Washington}
}

\author{Lin Lu}
\email{llu@sdu.edu.cn}
\affiliation{%
  \institution{Shandong University}
}

\begin{abstract}
\revision{
%disadvantages should be not only about surface quality but also printing efficiency.
%Inertia of the extruded material may damage the surface quality during transfer moves. The viscosity also makes support material hard to remove. 
%
We propose a novel computational framework for optimizing the toolpath continuity in fabricating surface models on an extrusion-based 3D printer. %\hs{Why specific Cartesian printer? We can not work on Delta/Polar/Robot arm based FDM printers?} 
Toolpath continuity has been a critical issue for extrusion-based fabrications that affects both quality and efficiency.
Transfer moves cause non-smooth or bumpy surfaces and get worse for materials with large inertia like clay.
For surface models, the effects of continuity are even more severe, in terms of surface quality and model stability.
%that both sides are of visual significance, making it impossible to hide any intermediate structures in the interiors. 
In this paper, we introduce an original criterion ``one-path-patch'' (OPP), for representing a shell surface patch that can be traversed in one path considering fabrication constraints.
We study the properties of an OPP and the merging operations for OPPs, and propose a bottom-up OPP merging procedure for decomposing the given shell surface into a minimal number of OPPs and generating the "as-continuous-as-possible" (ACAP) toolpath.
Furthermore, we customize the path planning algorithm with a curved layer printing scheme, which reduces the staircase defect and improves the toolpath continuity via possibly connecting multiple segments. 
%\hs{Missing the insight of including curved layers. Why do we include curved layers? Why curved layers is benefit to improve the continuity? Curved layers are easy understood to be useful for improve staircase. Why continuity?}
We evaluate the ACAP algorithm for both ceramic and thermoplastic materials, and results demonstrate that it improves the fabrication of surface models in both surface quality and efficiency. 
%These challenges even increase for thin shell surfaces, as both sides are of visual significance, making it impossible to hide any intermediate structures in the interiors. 
%Moreover, multiple open segments in one layer render the existing continuous toolpath algorithm infeasible\hs{too details...}.
%[3.Existing works has started to pay attention to this specific constraint;However, none existing works on shell models]
%\hs{We are the first including curved layers for ceramics printings?}
%[4.Our key technique framework]
%Then we introduce an original criterion ``one-path-patch'' (OPP), for representing a shell surface patch that can be traversed in one path in the context of curved layer printing considering fabrication constraints.
%, considering fabrication constraints \hs{move "considering fabrication constraints" to the last sentence? Because our OPP definition is constraint-aware.}.
%To conquer these challenges, we adopt a curved layer scheme for ceramics printing. 
%a decoupled orientation and support structures computation method.
}
\end{abstract}

%
% The code below should be generated by the tool at
% http://dl.acm.org/ccs.cfm
% Please copy and paste the code instead of the example below. 
%
\begin{CCSXML}
<ccs2012>
<concept>
<concept_id>10010147.10010371.10010396</concept_id>
<concept_desc>Computing methodologies~Shape modeling</concept_desc>
<concept_significance>500</concept_significance>
</concept>
<concept>
<concept_id>10010147.10010371.10010387</concept_id>
<concept_desc>Computing methodologies~Graphics systems and interfaces</concept_desc>
<concept_significance>300</concept_significance>
</concept>
</ccs2012>
\end{CCSXML}

\ccsdesc[500]{Computing methodologies~Shape modeling}
\ccsdesc[300]{Computing methodologies~Graphics systems and interfaces}

\acmJournal{TOG}

\keywords{Toolpath planning, shell models, extrusion-based printing}

%Fig. 1. Our system jointly explores the space of discrete design variants and fabrication plans to generate a Pareto front of (design, fabrication plan) pairs that minimize fabrication cost. In this figure, (a) is the input design for a chair and the Pareto front that only explores the space of fabrication plans for this design, (b) shows the Pareto front generated by joint exploration of both the design variants and fabrication plans for the chair, where each point is a (design, fabrication plan) pair. Design variations indicate different ways to compose the same 3D model from a collection of parts and are illustrated with the same color in the Pareto front. A physical chair is fabricated by following the result fabrication plan. This example shows that the fabrication cost can be significantly improved by exploring design variations.

%Ultimaker Cura

%Our framework greatly improves printed model quality by reducing the number of path disconnections as much as possible. Compared to the open-source state-of-the-art slicer cura (left) which generated void travels 195 times (each color represents a continuous path, and the red line represents void travels), our framework (right) divided model into 4 continuous paths, the printing quality is significantly improved and all paths satisfy manufacturing constraints.

\begin{teaserfigure}
%\vspace{-5pt}
  \includegraphics[width=\textwidth]{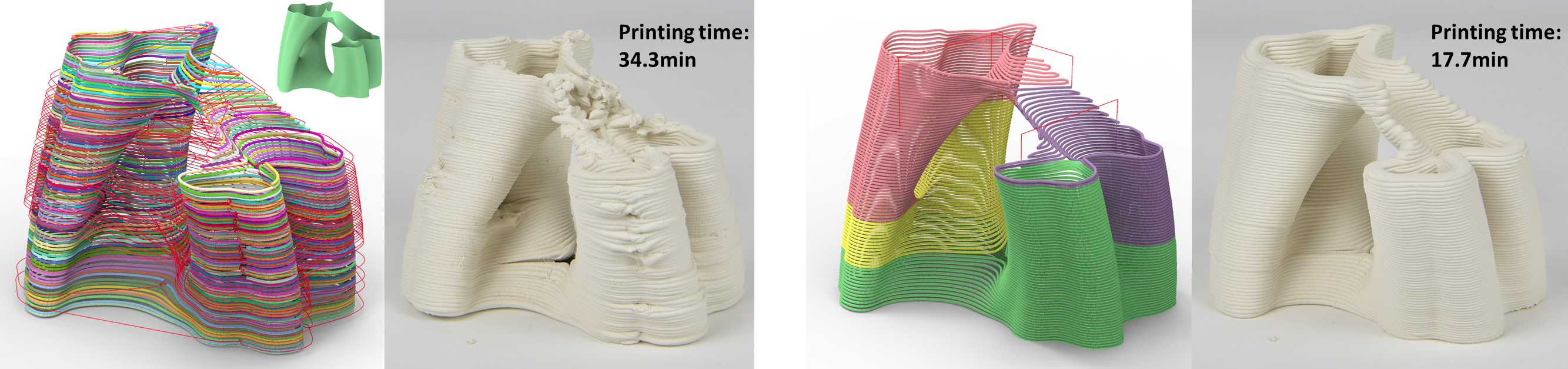}
	\vspace{-15pt}
  \caption{Our framework significantly improves fabrication efficiency and product quality of \revision{surface} models with \textit{as-continuous-as-possible} printing toolpaths. A novel geometric criterion ``one-path-patch'' (OPP) is proposed to decompose the input \revision{surface} into a minimal number of continuous printing patches with flat and curved \revision{collision-free} paths. 
  Compared with \revision{832} disconnected printing toolpaths with only flat slicing layers (left) generated by Ultimaker Cura software (different colors indicate each connected path, red lines represent the transfer moves), our framework produces 4 continuous deposition paths together with flat and curved slicing layers, realizing much better printing quality (right) and extremely efficient fabrication process (34.3 mins vs 17.7 mins).}
  \label{fig:teaser}    
\end{teaserfigure}
\newcommand{\hsyntax}[1]{\ensuremath{\mathit{#1}}}
\renewcommand{\grammarlabel}[2]{#1\hfill#2}

\newcommand{\egraph}{{e-graph}\xspace}
\newcommand{\eclass}{{e-class}\xspace}
\newcommand{\enode}{{e-node}\xspace}
\newcommand{\llhelm}{{LL-HELM}\xspace}
\newcommand{\hlhelm}{{HL-HELM}\xspace}
\newcommand{\submodule}{{e-class}\xspace}
\newcommand{\submodules}{{e-classes}\xspace}
\newcommand{\subprogram}{{sub-program}\xspace}
\newcommand{\subprograms}{{sub-programs}\xspace}

\newcommand{\tfas}{{target flat areas}\xspace}
\newcommand{\tfa}{{target flat areas}\xspace}
\maketitle

\section{Introduction}
\label{sec:intro}

% 1. continuity is a critical problem in fabrication 
% 2. shell is very popular (denoted as shell models)
% 3. continuity for shell models are even more critical 

% useful and general % Custom % 3-axis printing;
% Clarify that multiple-axis printing is...

%Continuous is important
%current continuity methods deal with closed surfaces
%how classic methods optimize the continuity? 

%\hs{reorganize introduction based on abstract. surface model, continuity (ceramic printing), existing works, propose OPP, curved layers, summary contribution}

%Reinforcement of General Shell Structures ~\cite{GilUreta2020}.

\revision{
Surface or shell models are widely used in structural design, being efficient in shape presentation and possessing featured functionalities like lightweight and effective thermal conductivity.
Potteries in the shell form have been developed since the Stone Age. 
In the context of additive manufacturing (AM), shells are cost-effective for both materials and fabrication time compared to solid ones.
In this paper, we focus on the surface model, particularly a thin shell with the thickness of a single path. It can be either open or closed, and therefore segments exist in layers for open surface models. 
}

\revision{
Continuity of the toolpath is one of the most fundamental problems in material extrusion-based AM.
Continuity of the nozzle's movement and material extrusion directly affects the surface quality, model stability, and fabrication efficiency.
For extrusion fabrication of surface models, the toolpath continuity plays a more critical role than solid models. 
Besides the surface quality and printing efficiency, transfer moves induce extra forces to the printed shell surface, weakening the stability, and the model may sag or collapse along with the accumulation of the forces.
}

%\hs{It seems that this paragraph comes too early?}
Especially, fabricating surface models with clay is getting popular thanks to the rapid progress on the ceramics printing techniques. The most feasible and cost-effective technique for 3D printing clay is direct ink writing (DIW), which shares the same architecture with fused deposition modeling (FDM), but has a larger opening nozzle that provides more material extrusion efficiency for high viscosity clay than thermoplastics~\cite{Chen2019}.  
%A typical example that combines the advantages of extrusion-based fabrication  and surface models is ceramic 3D printing with clay.
As a natural material, clay is environmentally friendly and durable.
%Ceramics printing has been gaining both industrial and academic interests in recent years.
%; thus, ceramic products are ubiquitous in construction, housing, consumer goods, etc.
Since the deposition rate of semi-liquid pastes is large, surface models are the most popular 3D printed objects.
Nevertheless, the artifacts on the surface quality caused by transfer moves cannot be negligible.

Existing path planning methods can be grouped into two broad categories.
One group focuses on optimizing the infilling patterns in each layer section~\cite{Zhao2016,Zhai2019} and achieves the continuous toolpath even in the layer with complicated contours; however, surface models with no interiors cannot take the advantage.
The other group is applicable to surface models, optimizing the sequences of contours~\cite{Lensgraf2017,Lensgraf2018,Yoo2020} based on search algorithms to achieve the minimal "extrusionless travel distance." 
Nevertheless, this wasted motion criterion is not equivalent to the toolpath continuity in terms of the number of transfer moves. The optimal continuous toolpath of a surface model is essentially a tailored surface decomposition problem considering the fabrication constraints.

\begin{figure}[t]
\centering
\includegraphics[width=1.0\linewidth]{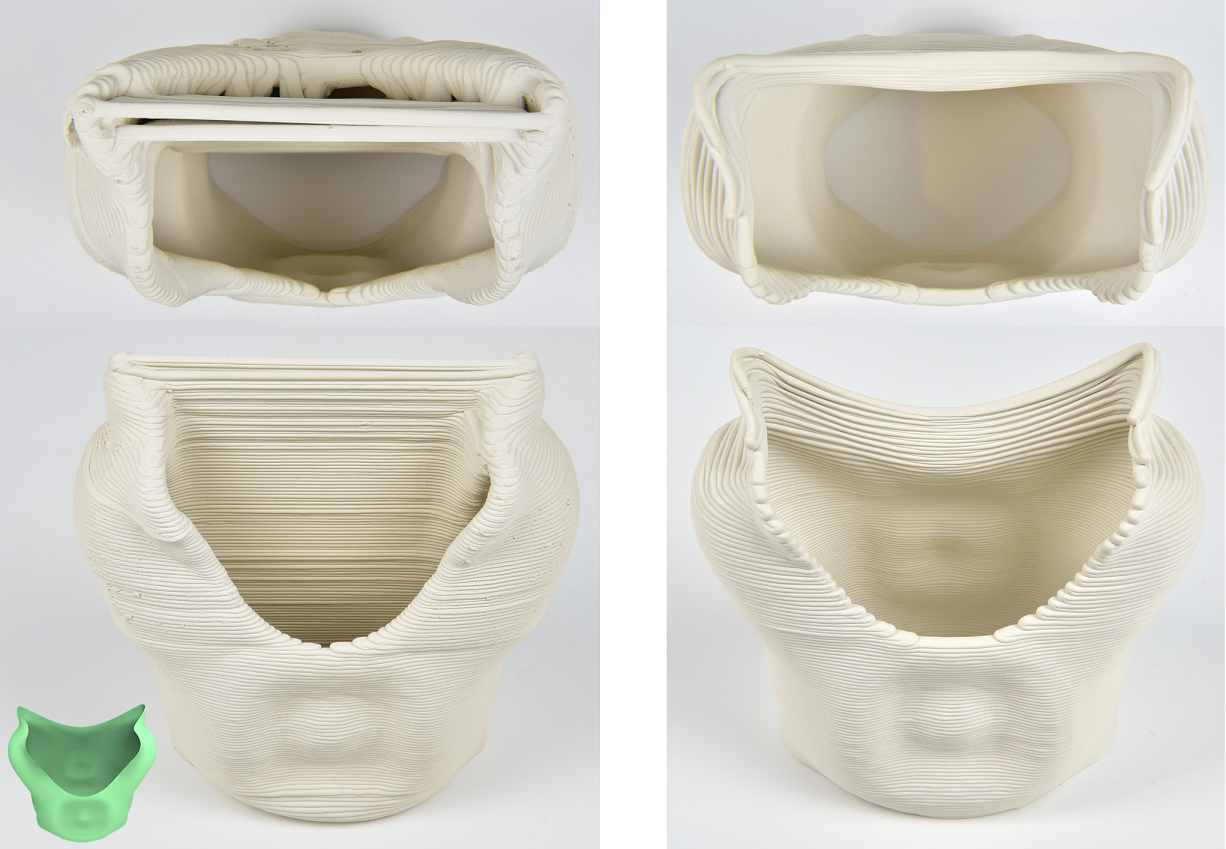}
\caption{Left: strictly-continuous path generated by \cite{Hergel2019}.  
\revision{
We tailor this method to surface models by slightly thickening the original surface into a valid volume. But the redundant non-model structure is still needed to connect separate model components (see the top view).} Right: our toolpath is fully continuous using curved layers without extra path.}
\label{fig:introcompare}
\end{figure}

%We mention that the continuity of ceramic 3D printing with clay is even more crucial due to a large amount of semi-liquid pastes during deposition and thus the artifacts caused by transfer moves.
Hergel~\etal~\shortcite{Hergel2019} recently proposed a path planning method for extrusion-based ceramics printing, which produces strictly continuous deposition paths that eliminate transfer moves. 
\revision{
This method works well with a single contour for each layer but cannot print multiple components per layer without adding non-model structures. 
Therefore, it cannot be easily adopted for shell models, as "hiding" the non-model intermediate structures without affecting the surface appearance is almost impossible, especially for open surfaces, referring to~\autoref{fig:introcompare}.
}
% closed surfaces, i.e., contours in each layer are all closed, such that bridges or interior support structures are "hidden" inside the shape. However, it 
% There is no interior or exterior for shell models, and both sides are of visual significance, making it impossible to hide any 
% Besides, multiple curvy segments exist in one layer. Open curves have two endpoints and thus increase double contact points between the non-model structure and model structure, 

%\hs{cite \cite{Li2021} in this paragraph?}

%As the most popular clay expression form, shell models like potteries \revision{have been developed} since the Stone Age. 
%Possessing featured functionalities like lightweight and effective thermal conductivity, shell models are widely used in oil and gas, aerospace, and craft industries~\cite{Bhatt2020}.
%In extrusion-based 3D printing, shell models are of high fabrication efficiency compared to solid ones. 

\revision{
Curved layers manufacturing is regarded as an effective method for removing the staircase defects~\cite{Etienne2019} and improving the strength through aligning filaments along the directions with large stresses~\cite{Fang2020}.
For surface models, curved layers have unique advantages that multiple components may be printed in a connected toolpath.
}
%It has been adapted to multi-axis printing platforms to strengthen the toolpath continuity for clay or concrete; however, it has never worked on the standard 3-axis platforms.

Adopting the curved layers scheme for Cartesian 3D printing faces two more constraints from fabrication besides continuity.
First, the layer thickness is adjustable but bounded by the extrusion amount. 
Second, the slope of the curved toolpath cannot be too steep, as the collision between the nozzle and printed model must be considered.
\revision{Considering the fabrication constraints, the problem can be regarded as a \textit{Precedence Constrained Minimum Path Cover Problem (PC-MPC)} }, which is equivalent to a classical NP-hard problem PC-TSP (see more details in~\autoref{conversion}).

%\begin{figure}[t]   
%\centering
%\includegraphics[width=1.0\linewidth]{images/feasible direction.png}
%\caption{Feasible directions for sample points on the shell model, sample points and feasible directions form a directed acyclic graph.}
%\label{fig:}
%\end{figure}

%\yl{This is a NP-hard problem, which can be proved by converting it into other classic NP-hard problem called precedence constrained traveling salesperson problem (PC-TSP). This process can be found in~\autoref{conversion}.}

%Restrict the printing to flat layers, this problem can also be expressed PC-TSP. Efforts in the literature have been made to approximated solutions on solving this problem to minimize the wasted motion and thus the overall fabrication time~\cite{Lensgraf2017,Lensgraf2018,Yoo2020}.

We target producing the ``as-continuous-as-possible'' (ACAP) toolpath for surface models, such that the number of transfer moves is minimized.
Our key idea is to propose a ``one-path-patch'' (OPP) criterion to represent a surface patch that can be printed in a continuous toolpath, in combined flat and curved layers.
We propose a bottom-up OPP merging algorithm for decomposing the given shell model
into a minimal number of OPPs and generate the ACAP toolpath.
This paper makes the following contributions:
\begin{itemize}

\item \revision{we introduce an original, fabrication constraints-aware, criterion  ``one-path-patch'' (OPP)  for representing a shell surface patch that can be printed in one path in the context of both flat layer and curved layer printing scheme.}

\item \revision{we propose a novel algorithm for decomposing the given shell surface into a minimal number of OPPs and generating the ``as-continuous-as-possible'' (ACAP) collision-free toolpath.}

%\item \revision{We provide a novel, computational framework for printing shell models on a standard 3-axis extrusion-based printing platform.}
\item \revision{we adapt our technique as a general computation framework for printing shell models in a ``as-continuous-as-possible'' manner on 3-axis extrusion-based printing platforms. 
}

% we adapt the centroidal Voronoi as a generative model for
%’like-natural’ honeycomb interior structures. This leads to a high strength-to-weight ratio.

\end{itemize}

\section{Related Work}
\label{sec:related}

\begin{figure*}[t]
\includegraphics[width=\linewidth]{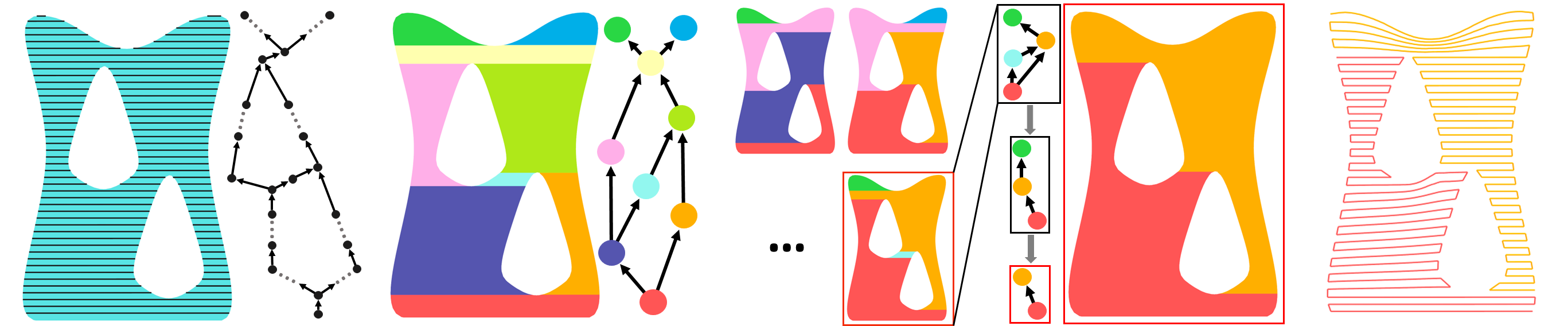}
\leftline{ \footnotesize  \hspace{0.033\linewidth}
            (a) Dependency \revision{relationship} \hspace{0.068\linewidth}
            (b) Initial \revision{decomposition} \hspace{0.044\linewidth}
            (c) \revision{Merging with flat layers}\hspace{0.026\linewidth}
            (d) \revision{Merging with curved layers}\hspace{0.027\linewidth}
            (e) \revision{Toolpath generation}}
\caption{
Overview of the ACAP algorithm. 
Given a surface model \revision{and a feasible orientation,} we slice horizontally along the Z-axis, \revision{and indicate the dependency relationship with a graph} (a) where each node represents a $segment$ or $contour$, and each edge represents the dependency of two nodes. 
\revision{The input model is initially decomposed, and each decomposed patch can be printed continually, whose dependency relationships are shown in} (b). Next, we \revision{merge small patches into bigger single printable ones, which may produce many multiple optimal merging solutions} (c). \revision{We further reduce the number of patches for each solution by applying curved layers} (d).
\revision{For the optimal merging solution with the least number of patches, we generate the toolpath for each patch and assign the printing order between them} (e).}
\label{fig:overview}
\end{figure*}

% Curved Layer Fused Deposition Modeling (CLFDM)~\cite{Chakraborty2008} allows continuous change of the z-value within individual layers, improving surface quality and strength of parts.
% The proposed method would be particularly advantageous over FDM in the manufacturing of thin, curved parts (shells) by reduction of stair-step effect, increase in strength and reduction in the number of layers.

\paragraph{Slicing and Path Planning}
In AM, the slicer is used for converting a model to toolpaths. Existing slicing algorithms divide the model as a stack of flat layers by geometric operations.
\cite{Lensgraf2016,Lensgraf2017,Lensgraf2018,Yoo2020} proposed a series of optimization algorithms to minimize the total
\textit{extrusionless travel distance} (wasted motion or print time) in the space of feasible toolpaths.
We represent the precedence constraints as the dependency graph in a similar manner.  
In contrast, we define the OPP criterion, reform the optimization in a more compact dependency graph, and gain a more efficient optimization framework. For most cases, we can achieve the optimal results rather than approximated ones.
%
% \hs{why do we propose agreed method for flat layer merging ? not use existing methods?1: optimal for less points; 2:output multiple solution; 3:0/1,weight is not a float.}
Many efforts have been made to optimize the toolpath in terms of continuity, filling rates, and mechanical properties~\cite{Zhao2016, Zhai2019,Xia2020}.
\revision{
Adaptive bead width control could further reduce the 
under- and over-filling artifacts~\cite{Kuipers2020,Hornus2020}. 
}
%\cite{Hornus2020} propose a variable-width contouring technique that further optimizes the filling density.
%
Hergel~\etal\shortcite{Hergel2019} present a method for generating strictly-continuous, self-supporting deposition paths for extrusion-based ceramic printing, performing nicely on watertight geometric models. However, this method cannot directly apply to shell models as illustrated in \autoref{fig:introcompare}. 
%Applying their method in thin-shell models makes it more challenging to remove support structures. Considering that thin shell models is more fragile than solid, manual removal will significantly damage the surface quality of the model and even cause breakage.
%As said in their paper, there is a trade-off between leaving support
%removal marks and accepting the artifacts that would result from transfer moves.
%\cite{Zhao2016} develop a space filling curve called \textit{connected Fermat spirals}, the toolpath is continuous and low-curvature. 

\paragraph{Curved Layers Printing}

Compared with traditional flat layers, curved layers contain dynamic z-values within individual layers and have excellent properties in AM. Such as alleviating the staircase defect, improving surface smoothness, strengthening the printing model and reducing printing time. 
The first discussion about curved layers is called curved layers fused deposition modeling (CLFDM)~\cite{Chakraborty2008}, which allows continuous change of the z-value within individual layers. 
Afterwards, \cite{Huang2012,Allen2015,Llewellyn-Jones2016} perform physical experiments with FDM printers to demonstrate these properties.  
%\cite{Allen2015} and \cite{Llewellyn-Jones2016} mix flat layers and curved layers. Flat layers print the interior of the model, and curved layers print the thin skin surface. They did physical experiments with the delta-style FFF printer to demonstrate improving mechanical properties and surface finish. 
\revision{The industry standard slicing software Ultimaker Cura~\cite{Cura} also involves the curved layer printing in the surface mode that produces spiralized outer contours of the mesh instead of the solid model, which works well on simple shapes like a vase or cylinder.
}
\cite{Ezair2018} present an algorithm that generates covering curves based on geometric characteristics of a given volume. 
\cite{Etienne2019} take a different approach that optimizes a parameterization to obtain smooth surface tops.
The produced toolpaths are mapped back into the initial domain without requiring splitting or re-ordering. 
We apply this method in our curved OPP merging.
% takes a different point of view to generate curved layers. Firstly, they deform the input model by optimizing a mapping with fabrication constraints. Then use flat layers to slice the deformed model. Finally, flat layers are mapped back into the original space to generate curved layers. \hs{maybe just list a basic idea of their method. Then give them a credit to show how we apply their method in this paper.}
%The above work makes full use of the three DOFs of traditional printers. However, the normal of the printing path is not align with the printer-head, local gouging can easily occur in steep slope~\cite{Chen2019RCIM}. 
%
Recently, researchers have applied curved layers printing on multi-axis printers. With the help of additional DOF, \cite{Dai2018,Xu2019,Li2021} design curved toolpath to fabricate solid models in a support-free way. 
\cite{Chen2019RCIM} present a new CLFDM slicing algorithm that allows variable thickness layers.
\cite{Fang2020} introduce a field-based optimization framework to generate curved layers for reinforcing the mechanical strength of 3D printed models. %\llu{not very good}
However, we focus on three-axis printer platforms in this paper.
%However, multi-axis printing is not friendly to clay, as it does not dry instantly and might collapse when the orientation changes.  \llu{here need some revision}
%In our method, we print in 3-axis and consider fabrication constraints with curved layers.

\paragraph{Fabrication of Thin Shells}
Fabricating thin shells is gaining increasing attention as it accelerates the fabrication time compared to closed models and lightweight shell models are of wide applications. 
This advantage is further enhanced when fabricating viscous slurry materials, like clay and concrete, with large extrusion amount and fast deposition rate.
For shell models, the continuity of material deposition is more critical due to the apparent artifacts caused by transfer moves. Most works in the literature adopt multi-axis platforms and incorporate curved layers for printing shells. 
%To guarantee the continuity of material deposition, research works also choose to bypass the conventional slicing step. 
\cite{Mitropoulou2020} present a method to design non-planar layered print paths for robotic FDM printing of single-shell surfaces. 
\cite{Bhatt2020} propose the layer slicing and toolpath planning algorithm to build 
 thin shell parts on a 3-DOF build-platform and a 3-DOF extrusion tool.
Printing concrete shells is attracting interests in the interdisciplinary area of digital fabrication and architecture. 
\cite{Burger2020} use the single-shell as molds for concrete casting. 
\cite{Anton2019} propose a design tool for producing 
bespoke concrete columns and involve curved layer for continuous extrusion.
\cite{BHOOSHAN2020} also emphasize on interactively shell modeling and integrate modeling with toolpath generation.

\paragraph{Decomposition for Fabrication}
Many efforts have been focused on model decomposition for fabrication. 
Objectives for model decomposition include fabricating the model that satisfy the constraint, improving surface quality, saving or avoiding support structures, and reducing printing time. 
\cite{Luo2012} propose a solution to decompose the model into smaller parts that every part can fit the printing platform. In addition to considering the criterion of printing volume, structural soundness and aesthetics are their decomposition objectives. 
For improving surface quality, \cite{Hildebrand2013} generate a partition and compute the optimal slicing direction for subparts.
\cite{Hu2014} decompose a given shape into as few approximate pyramidal parts as possible. Their motivation is that pyramidal shapes are well suitable for fabrication.
\cite{Vanek2014,Wei2018} decompose shell models into small parts to save the support material and reduce the printing time. A manual assembly process is required after printing all shells. 
Also, to avoid supporting materials, \cite{Wu2017, Wu2020}'s decomposition approaches consider the collision-free constraint and sequence of printing. They printed models in a multi-DOF 3D printing system so that manual assembly is not required. 
\cite{Herholz2015,Muntoni2018} decompose general three-dimensional geometries to satisfy the height field constraint. 
To minimize the number of cutter setups for finish-stage machining in CNC, \cite{Zhao2018} develop an algorithm to perform surface decomposition with the accessibility constraint. 
The above methods did not consider the continuity of the printing path as a criterion for model decomposition. 
It is worth mentioning that \cite{mahdavi2020vdac} propose carvability criteria for continually carving a connected domain, which requires both visibility and monotonicity. However, they did not take curved slicing layers into account. In such situation, the key difference between our OPP criterion and carvability is that OPP does not require visibility but support curved layering fabrication.

\paragraph{Ceramic Printing}
Ceramic materials in AM have attracted heightened attention in recent years~\cite{Zocca2015, Chen2019}. 
It would reduce both processes and resources required to produce geometrically complex shapes in the traditional ceramics industry, and thus nurture new ideas or applications in architectural decorations~\cite{Chan2020}, arts, etc. Researchers are also developing advanced engineering ceramics, such as metal oxides, carbides, and nitrides, to specific engineering demands~\cite{Peng2019}. 
Existing work mainly focuses on studying formulations of the water-to-clay ratios and some additives, and physical analysis of the sintered models in terms of compression, thermal stability, etc.~\cite{Revelo2018, Ordonez2019}.

Even though the slicing and toolpath planning for DIW ceramic printing share both constraints and objectives with FDM, it possesses additional constraints due to the viscosity of clay, which is attracting the attention of researchers. 
Recent attempts consider the path planning for closed model~\cite{Hergel2019}, integrated modeling and path generation for relatively simple shapes~\cite{Zhong2020}, and stability enhancement for shell models~\cite{Xing2021}. 
While for general shell models, there are still no effective path planning methods.

%\hs{Indeed, our VDAC paper considering toolpath continuity during decomposition. However, we did not refer that paper. So it seems fine.}.

%\cite{Vanek2014} proposed an optimization framework to decompose solid models which are hollowed its inner parts into thin shell surfaces for saving printing time and the support material. There is a manual assembly process after printing all shells.

\section{Overview}
\label{sec:overview}

%considering flatten area and support structure volume.

\revision{Given a thin shell model $M$ with a feasible orientation that meets the support structure constraint, our algorithm} aims to achieve maximal path continuity, i.e., decompose $M$ into \revision{a minimum number of surface patches where each patch can be printed consecutively. For narrative convenience, in this paper we define a printable surface patch as a patch that can be printed with a single path.} 
%considering the path layer thickness constraint and slope angle constraint (\autoref{sec:fabcon}). The formulation of OPP is detailed in~\autoref{sec:oppdef}.

\revision{
As a PC-MPC problem, such decomposition is NP-hard that no efficient solution on all possible inputs. 
Our key idea is to apply an over-segmentation followed by a bottom-up merging procedure.
We first slice $M$ through uniformly distributed flat planers (\autoref{sec:dependency-graph}). Each sliced element can be seen as a single printable surface patch. Multiple printing paths from mutually contiguous sliced elements can be connected into a single path by including a set of short connecting paths, which indicates that we can reduce the number of printable patches via merging small initial patches (\autoref{sec:I-OPP-Merging}). 
Besides, we observe that curved slicing layers can be exceptionally effective in generating continuous printing paths for multiple separate printable surface patches of flat layers. The number of printable patches can be further reduced by replacing flat layers with curved slicing layers as much as possible (\autoref{sec:combination}).}

\revision{
The merging criterion is the main challenge for the bottom-up merging and curved layers replacement process. We need to formulate the "sub-surface-patches" that can be merged into a single printable patch or replaced by curved layers. 
The new concept of "one-path-patch" (OPP) is introduced as the merging criterion (\autoref{sec:oppdef}).
As a final step, we carry out the path planning for each OPP, followed by a post-optimization process to improve the path smoothness and spacing (\autoref{sec:PathGeneration}). 
Note that we avoid the potential global collision between the printer and printed layers considering the nozzle size (\autoref{sec:extendToGeneralNozzle}).
Our algorithm pipeline is illustrated in~\autoref{fig:overview}.
%The pseudo-code is presented in the appendix.
} 

%We propose a bottom-up merging procedure to convert the PC-MPC problem into a much more tractable combinatorial optimization problem. 
%The basic idea is to apply the flat-layer based merging (\autoref{sec:dependency-graph},~\autoref{sec:I-OPP-Merging}) then the curved-layer based merging (\autoref{sec:combination}). Finally, carry out the path planning for each OPP, followed by a post-optimization process to improve the path smoothness and spacing (\autoref{sec:PathGeneration}). Our algorithm pipeline is illustrated in Figure~\ref{fig:overview}.

\subsection{Fabrication Constraints}
\label{sec:fabcon}
We mainly consider three fabrication constraints in this work. The first bounds a feasible range of layer thickness. The second is to avoid collision between the extrusion device (nozzle, extruder, and carriage) and the printed parts while operating toolpath from curved layers. The third describes the geometric requirement for the decoupling fabrication strategy of the intact model and support structures.

%: slope angle, thickness constraint.\hs{Are there any other constraints we ignored?}
%The first is to avoid local gouging. The second is related to feasible thickness. In addition, our algorithm has additional constraints on support.\hs{Then say we have three constraints.?}
%\begin{figure}[t]
%\centering
%\includegraphics[width=1.0\linewidth]{images/thickness constraint.png}
% \leftline{ \footnotesize  \hspace{0.15\linewidth}
%             {(a)} \hspace{0.28\linewidth}
%             (b)\hspace{0.3\linewidth}
%             (c)}
%\caption{Layer thickness constraint. Too small layer thickness, e.g. 0.2mm, results in over-stacking (left); and too large thickness, e.g. 3.5mm, causes under-stacking (right). The middle shows the thickness (1mm) within the constraints produces good quality results.}
%\label{fig:thickness constraint}
%\end{figure}

\paragraph{Thickness constraint}
\begin{wrapfigure}[12]{r}{0.09\textwidth}
\vspace{-12pt}
\hspace{-20pt}
\centering
\includegraphics[width=0.12\textwidth]{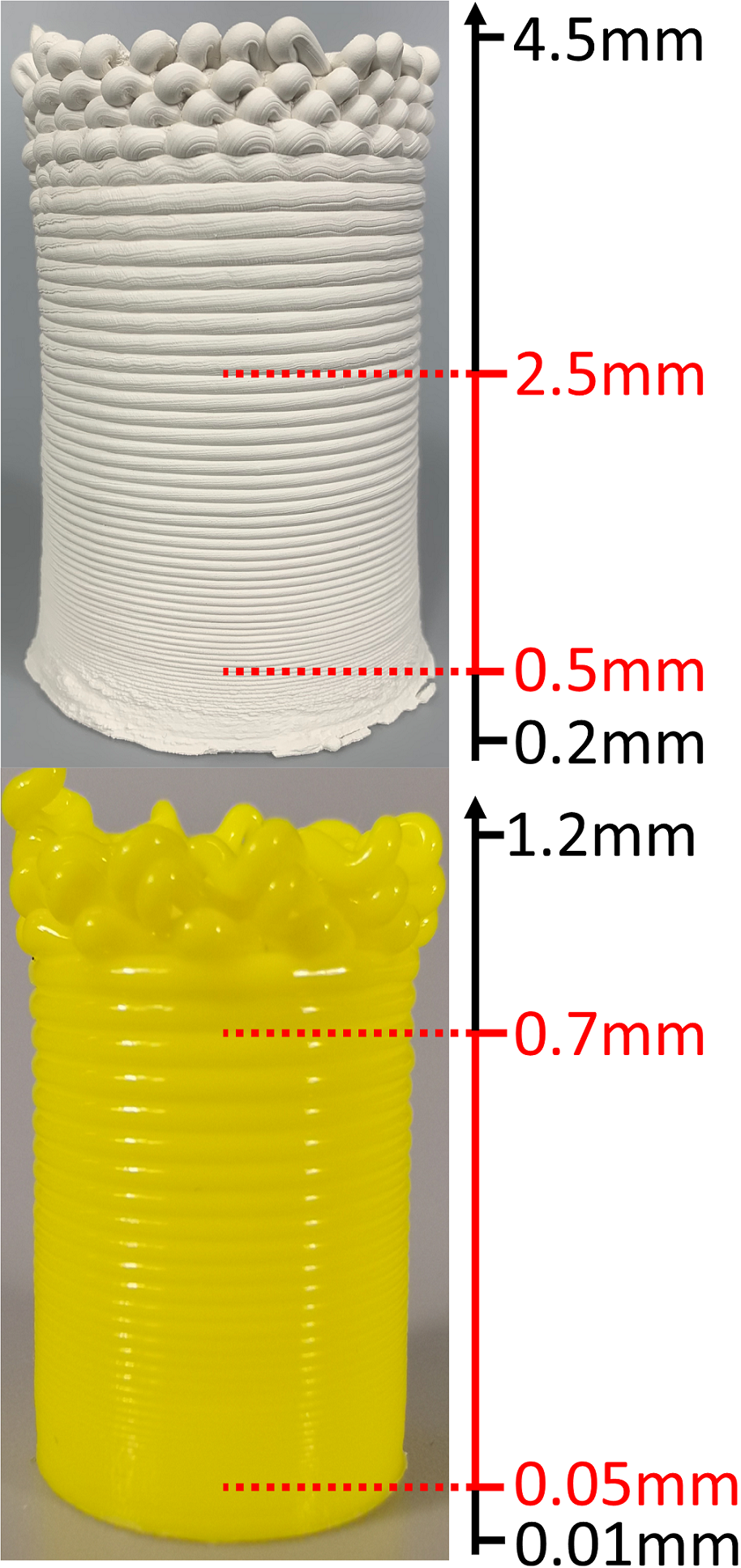}
\end{wrapfigure}

Affected by the fluidity of materials, too small layer thickness would cause the layers to squeeze together. Such an over-stacking effect results in artifacts on the surface (see inset). While too large thickness results in under-stacking, the adjacent layers are not well bonded. We use $t_{min}$ and $t_{max}$ to represent the minimum and maximum layer thickness. 
\revision{For ceramic printing, we take $t_{min}=0.5mm$ and $t_{max}=2.5mm$. For FDM printing, $t_{min}=0.05mm$ and $t_{max}=0.7mm$. 
This range is used as a constraint when generating the curved layers.}

\begin{figure}[t]
\centering
\includegraphics[width=1.0\linewidth]{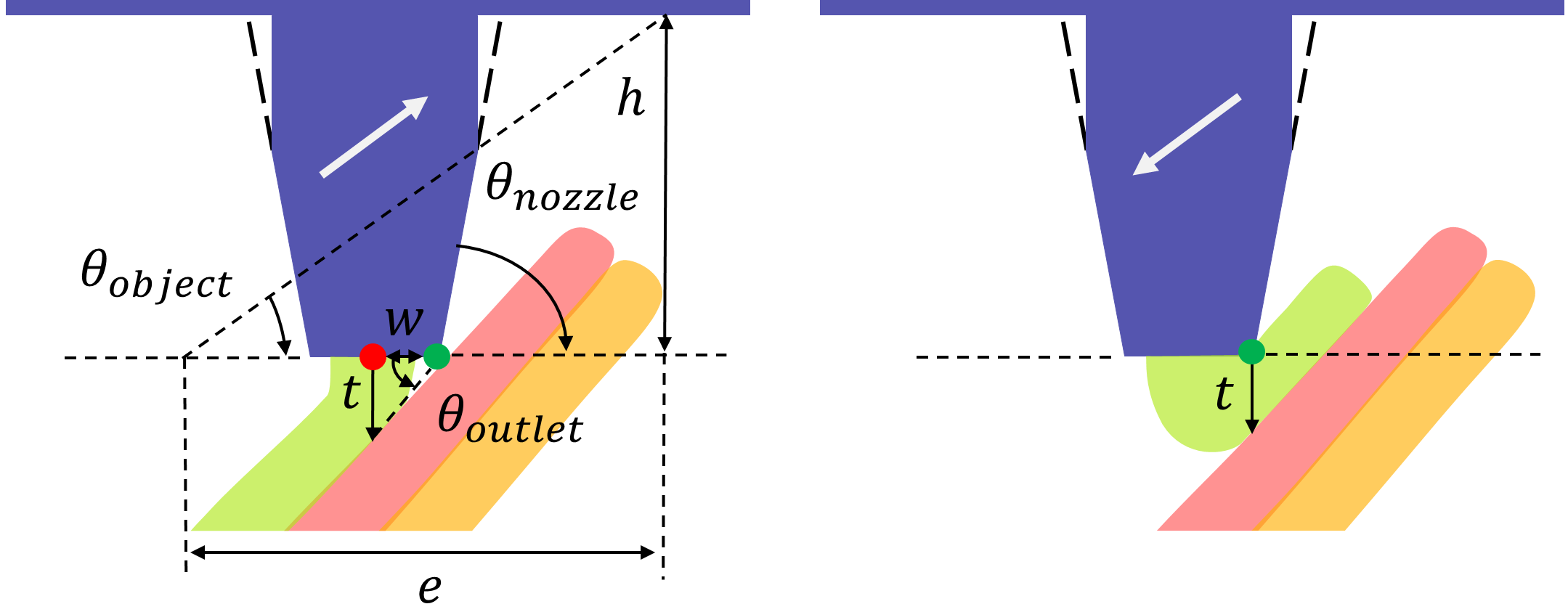}
\caption{\revision{Slope angle constraint of curved path.
The nozzle is a combination of a cylinder and a truncated cone (purple). We conservatively simplify it to a truncated cone to calculate the angles.} 
%$\theta_{nozzle}$ and $\theta_{object}$ are to avoid collision with the extrusion device.
$\theta_{nozzle}$ indicates the angle between the nozzle tip and the horizontal. $\theta_{object} = \tan^{-1}\frac{h}{e}$, where $h$ is the vertical distance between carriage and nozzle tip and $e$ is the maximum XY extent of the already printed object. $\theta_{outlet} = \tan^{-1}\frac{t}{w}$, where $w$ is \revision{the nozzle radius}. $t$ represents a reference layer thickness. (left) and (right) represent the nozzle going uphill and downhill, respectively.
}
%\caption{Slope angle constraint of curved path, the nozzle (orange) and carriage (gray) should avoid collision with the printed layers (blue) in movement. $t$ represents the maximum allowable layer thickness. $\theta_{nozzle}$ and $\theta_{carriage}$ are the angle between horizontal plane with outer wall of the nozzle and between horizontal plane with carriage, respectively, and $\theta_{outlet}$ is the angle between the previous printed layer and the horizontal plane that is needless when obliquely down printing. \hs{what's the blue line? What is the layer?} \hs{It seems you define these angles twice times (also in the text).} \hs{I still confuse why upward and down is different.} \hs{The visualization can be better.}}
\label{fig:slope constraint}
\end{figure}

\paragraph{Slope angle constraint}
\revision{
In \cite{Etienne2019}, the collision constraint has been modeled as an inverted cone to forbid already printed parts from entering. A local slope angle constraint for the printing paths is extracted from the forbidden cone as $\theta_{max} = \min(\theta_{nozzle}, \theta_{object})$ (see~\autoref{fig:slope constraint} for the detailed formulation).  Instead of regarding the printer as a pointed conical nozzle by overlooking the nozzle part's flat outlet, we propose making this formulation more precise. We represent the nozzle as a combination of a cylinder and a truncated cone.}

We make an interesting observation that the possible collision differently when going uphill and downhill. See \autoref{fig:slope constraint}, when going uphill (left), since the point on the right side of the outlet (green dot) is closest to the layer printed below, it may collide first, which can be avoided by restricting the path slope angle as $\theta_{outlet}$. Conversely, when going downhill (right), the nozzle has to be raised by $t$ to avoid self-layer collision between the point on the right side of the outlet (green dot) and the current layer. There is no need to define any slope angle for the downhill case with the raising operation. Finally, set the upper bound of the path slope angle as $\theta_{max} = \min(\theta_{nozzle},\theta_{object},\theta_{outlet})$. 
\revision{Note that this is a local constraint. The global collision caused by the height of the printed model exceeding the nozzle length is not considered. The solution will be mentioned in \autoref{sec:extendToGeneralNozzle}.}

\paragraph{Support structure constraint}
\revision{
We involve restrictions on the support structures for surface models, as support structures degrade the surface quality~\cite{Hergel2019}.
For surface models that are not self-supporting, we decouple the fabrication of the intact model and support structures, i.e., we pre-print the support structures and place them during the fabrication process.
This requires that the support structures locate on the ground and form a height field volume related to the printing orientation, as the support structures placed on the model may induce too much weight for the printed shell to afford.
For such models, we decompose them into multiple patches and assemble them after fabrication (an example is shown in \autoref{fig:kitten}).
}

% It's hard to place the membrane and remove the support structures in such a case. In addition, the external support structures may induce collapse of the part printed below while it has not been well solidified. 

%to make it easy to remove the support and eliminate the artifact caused by the support removal process, we decouple the fabrication for the intact model and support structures.
%Specifically, we add a membrane between the model and pre-fabricated supports during printing. Then, the support structure can be easily removed after clay materials drying. \hs{I remove this sentence which has been told in the introduction}
%
%A general fabrication process includes: 1) fabricate the support structures; 2) place a membrane over; 3) fabricate the intact model. %We remark that this decoupling does not work for support structures on the model instead of the ground. 
%
%Therefore, we only handle self-supported models or support structures falling on the ground (only need to add membrane once). 
\subsection{One-Path-Patch (OPP)}
\label{sec:oppdef}

%concept-level definition
%To maximize the continuity of ceramics printing for shell models, our technique aims to decompose the input shell model into minimal number of "one-path-patches" (OPPs). One OPP is a connected manifold surface patch which can be printed with a fully continuous toolpath under the fabrication constraints of ceramics printing, illustrated in~\autoref{fig:OPP-Demo}. 

\begin{figure}[t]   
\centering
\includegraphics[width=0.85\linewidth]{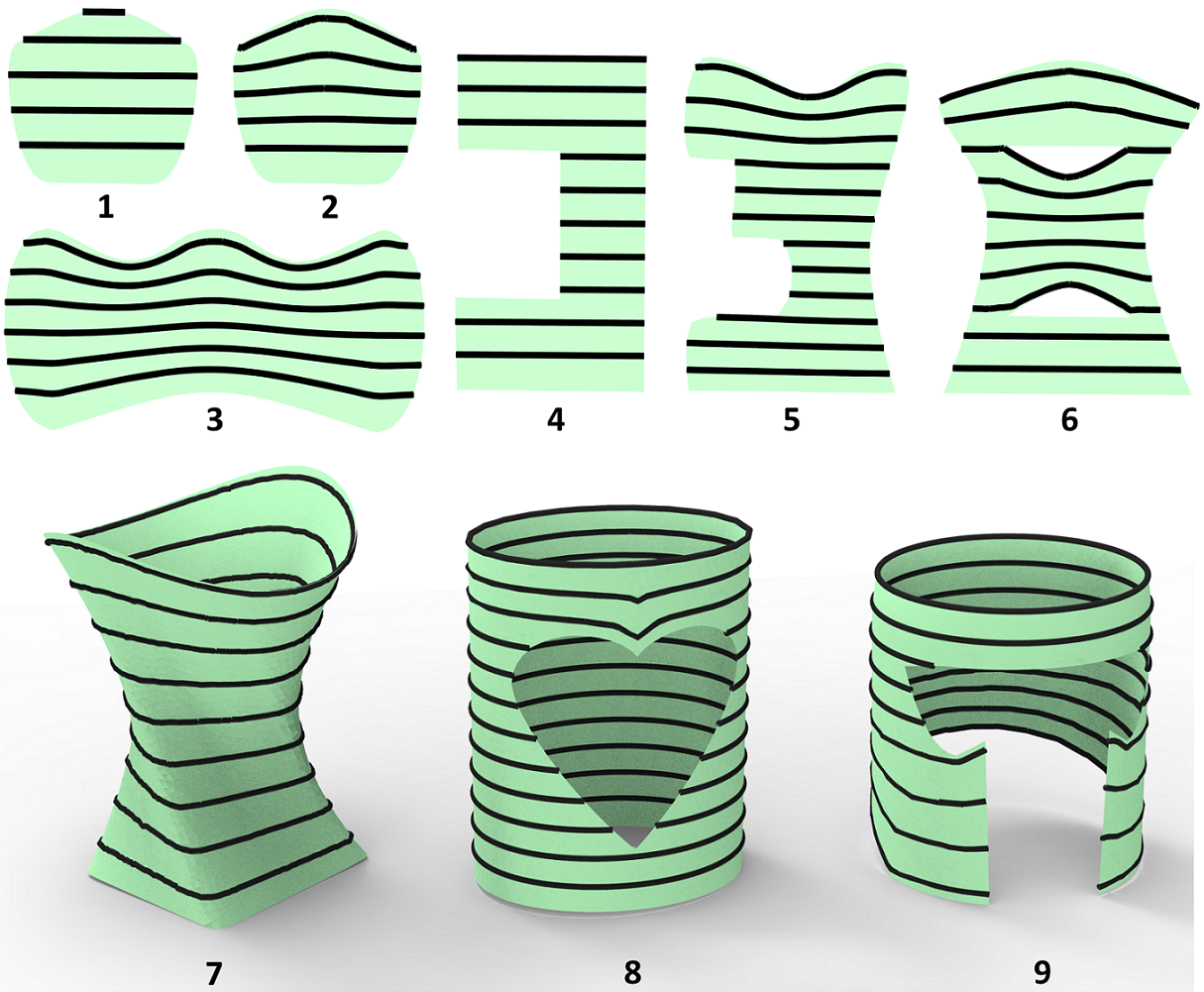}
%\vspace{-8pt}
\caption{
Illustration of OPPs that can be printed in a single toolpath generated from flat layers (1,4), curved layers (2,3,5,7), or combined flat and curved layers (6,8,9). One OPP can be sliced with different strategies (1,2).
%Illustration of an OPP which can be printed in a single toolpath with the curved slicing layers (a), not the planar slicing layers (b). The surface patches (c-d) are not OPP where (c) violates the slope angle constraint and (d) violates the thickness constraint.
%When using flat layers, a patch is divided two OPP (a), same patch can be divided only one OPP by using curved layers (b). The curved layers slope is too large (c) and thickness is too thin (d), which violate manufacturing constraints.
}
\label{fig:OPP-Demo}
\end{figure}

% recall definition of OPP;
\revision{
Recall that we aim to maximize the continuity by decomposing the input shell model into the minimal number of printable "one-path-patches" (OPPs). }
%{\color{blue}We first define the layer wasn't connected from end to end as \textit{segment}, otherwise as \textit{contour}.}\fc{we did not define segment and contour before?}
\revision{
With respect to a printing direction, a manifold surface patch is a printable OPP, iff 1) there exists a set of slicing layers where each layer orthogonal to the printing direction intersects the patch resulting in a single \textit{segment} or \textit{contour}; 2) the resulting intersected \textit{segments}/\textit{contours} satisfies the fabrication constraints. In the paper, we deliberately choose a low-resolution layer height to make the printing paths more visible.}
%As formulated in Section~\ref{sec:intro}, a manifold connected surface patch $P$ is an OPP associated with a set of slicing elements (\textit{segments} or \textit{contours}) where all intersected elements come from different slicing layers under the fabrication constraints.
Three types of OPP can be defined according to the slicing layers: (I) only flat layers; (II) only curved layers; (III) combination of I and II, named as I-OPP, II-OPP, and III-OPP for short, shown in~\autoref{fig:OPP-Demo}. Curved layers of OPPs would take the thickness constraint and slope angle constraint.

%~\cite{Etienne2019}
%We therefore initially attempt to flatten all surfaces that could
%be possibly reproduced under the maximum printable slope θmax.

\begin{figure}[t]   
\centering
\includegraphics[width=0.9\linewidth]{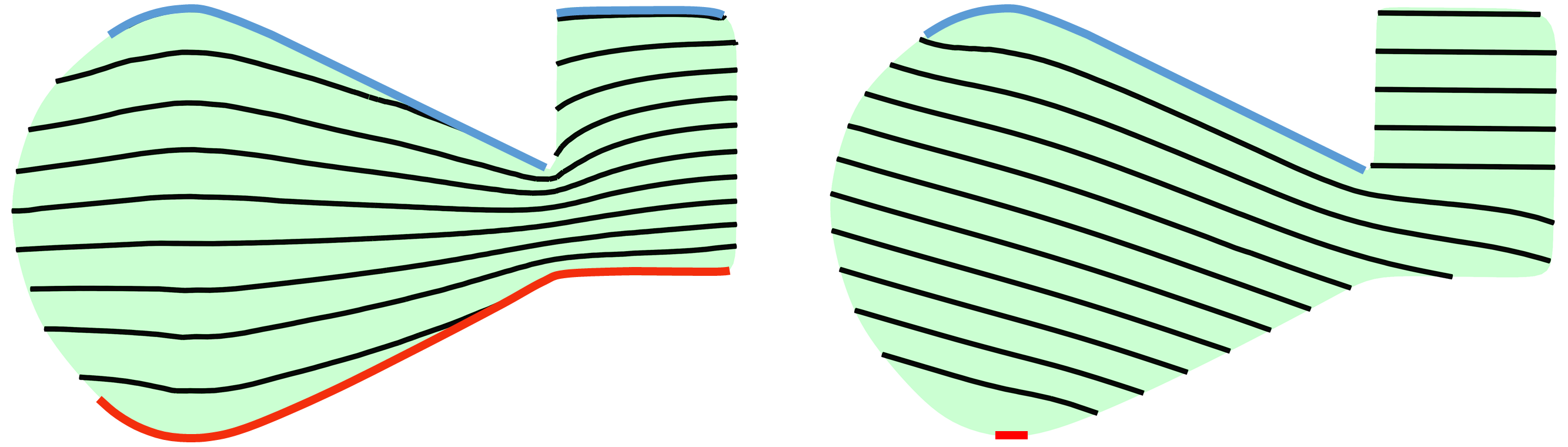}
%\vspace{-8pt}
\caption{
An illustration that staircase minimization is not equivalent to continuity maximization. Left visualizes the CurviSlicer result with the \tfas (initially attempt to flatten all areas under a specific slope angle, red segments). It could only flatten a part of the \tfas to avoid violating layer thickness constraint. The result layers cannot produce OPP layers.
Right shows the OPP layers and the specific \tfas of our method.
Note that different from CurviSlicer, the downward facing areas can also be taken as \tfas thanks to the support structure constraint (\autoref{sec:fabcon}).}
\label{fig:curvislicer}
\end{figure}

%extend the insight of the opp definition
% with only flat layers=>
What geometric properties should an OPP hold?
For type I, the OPP criterion is equivalent to its \textit{monotonicity}\footnote{A 2D polygon $P$ is \textit{monotone} with respect to a straight line $L$, if every line orthogonal to $L$, intersects $P$ at most twice. A 3D manifold surface patch is \textit{monotone} in direction $L$ if all cross-sections orthogonal to $L$ are single section~\cite{toussaint1985movable}.}, as (1,4) shown in~\autoref{fig:OPP-Demo}. 
%\hs{clarify the difference from carvability? Related works?}z
%compare with the carvability.
%Do we need to clarify the difference from carvability?
%It's worth to mention that Mahdavi-Amiri et al. proposed carvability criteria for continually carving a connected domain which requires both visibility and monotonicity ~\cite{mahdavi2020vdac}. They did not take curved slicing layers into account. In such situation, the key difference between our OPP criteria and carvability is that OPP does not require visibility.
As for type II, the intrinsic geometric properties of an OPP is hard to propose, where \textit{monotonicity} becomes a sufficient and unnecessary condition. In~\autoref{fig:OPP-Demo}, (3,5,7) is an OPP but not \textit{monotone}, and it demonstrates that a single OPP with curved layers could cover regions where multiple I-OPPs are applied.
%Benefit of curved layers

To determine whether $P$ is an II-OPP, one option can be to extract curved layers directly and then assess the two fabrication criteria.
CurviSlicer seems a perfect match for this, which desires to flatten as many areas as possible to minimize staircases~\cite{Etienne2019}. 
However, staircase minimization is not always equivalent to continuity maximization, which is sensitive to the \tfas taken as the input of CurviSlicer, as shown in \autoref{fig:curvislicer}.
Two I/II-OPPs can be merged to a single II-OPP (details in \autoref{sec:judgement}). For a whole shell model, a bottom-up OPP merging procedure is introduced to minimize the number of OPPs, during which process III-OPPs (as (6,8,9) in \autoref{fig:OPP-Demo}) are generated (\autoref{sec:combination}).

\begin{wrapfigure}[4]{r}{0.13\textwidth}
\vspace{-13pt}
\hspace{-23pt}
\centering
\includegraphics[width=0.15\textwidth]{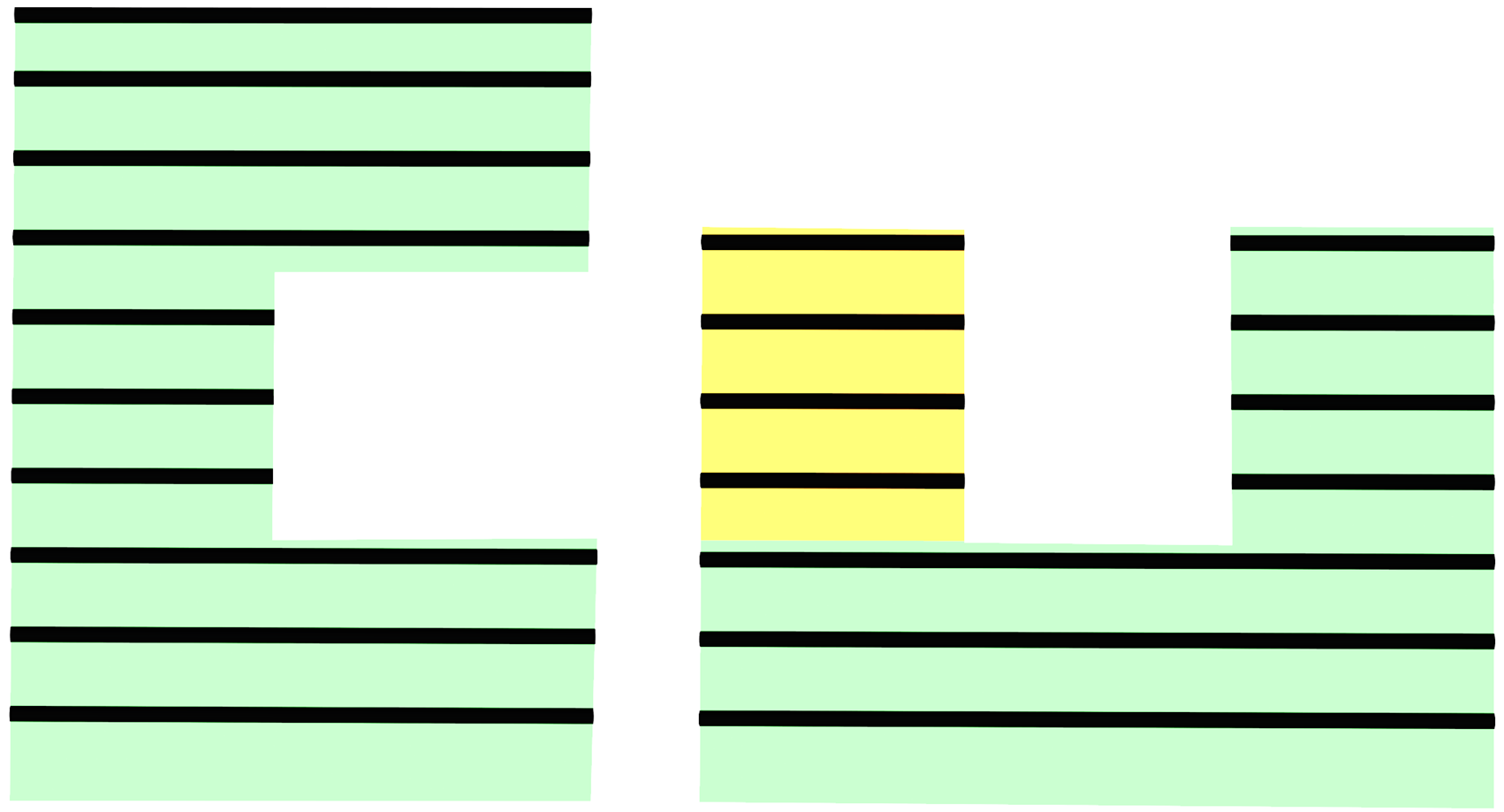}
\end{wrapfigure}

%However, since it doesn't care about the continuity of the path, it cannot guarantee the judgment is correct\hs{where is the example?}, and it takes a long time to make a judgment once because need to solve huge equations, most models\hs{what models?} require tens of thousands \hs{what unit?} of judgments, which is unacceptable in time. \hs{evidence is not enough.} 

\begin{figure}[t]   
\centering
\includegraphics[width=1.0\linewidth]{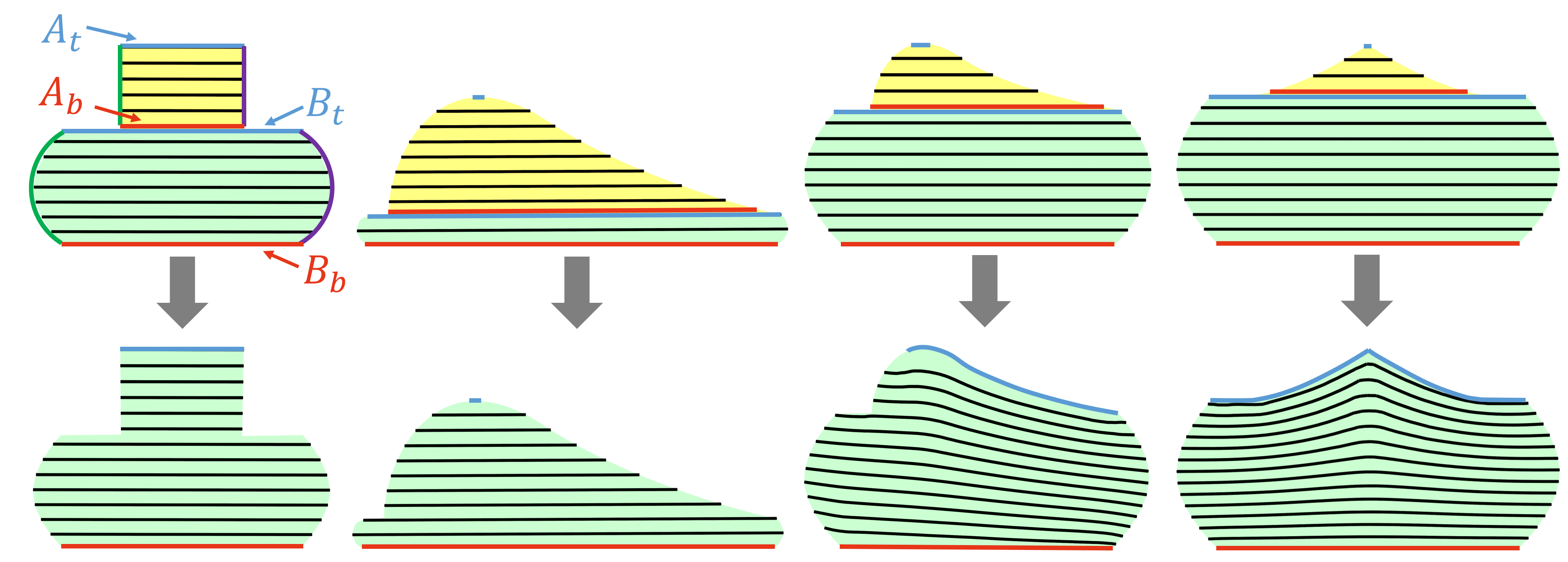}
%\vspace{-8pt}
\leftline{ \footnotesize  \hspace{0.075\linewidth}
            {(a)} \hspace{0.21\linewidth}
            (b)\hspace{0.22\linewidth}
            (c)\hspace{0.21\linewidth}
            (d)}
\caption{Illustration of merging two OPPs by \textit{stacking} operation (a)(b) and \textit{curving} operation (c)(d).
(a) and (b) cannot be merged via curved layers for violating the slope angle constraint and thickness constraint, respectively.  
\revision{
The red and blue segments indicate the top and bottom \tfas ($A_{t}, A_{b}, B_{t}, B_{b}$). The green and purple lines indicate two oblique polylines.}
}
\label{fig:OPPMerge}
\end{figure}

%properties of an OPP
To the best of our knowledge, the OPP criterion has never been explored before. An OPP possesses three key properties as the elementary path generation element to maximize the continuity of extrusion-based printing:  

\begin{enumerate}
%1：respect to a specific fabrication direction
\item The OPP criterion is defined with respect to a specific printing direction. An OPP in a printing direction (left) may not be an OPP with another direction (right), shown in the inset.
%2: the slicing layers are not unique
%~\autoref{fig:OPP-Demo}
\item Even with the same printing direction, valid slicing strategies of an OPP may not be unique. In~\autoref{fig:OPP-Demo}, (1) and (2) show two kinds of slicing layers where flat and curved slicing layers can be applied. Curved layers always produce fewer staircases associated with higher priority than flat layers.
%OPP union
\item Two OPPs can be merged to one single OPP. 
Two OPPs (A and B) can be merged via two operations, \textit{stacking} and \textit{curving}, as shown in~\autoref{fig:OPPMerge}. 
\textit{Stacking} indicates that 1) OPP A's bottom layer \revision{$A_{b}$} locates on the above neighboring layer of B's top layer \revision{$B_{t}$}; 2) \revision{$A_{b}$} and \revision{$B_{t}$} can be connected by its two end points.
\textit{Curving} indicates the merging operation of \autoref{sec:judgement}. The two operations are used in \autoref{sec:combination} to minimize the number of OPPs.
\end{enumerate}

\subsection{Curving Operation}
\label{sec:judgement}
%input
The input of this operation are two I/II-OPPs (A and B) that can be originally merged via \textit{stacking}. \textit{Curving} operation aims to output the merged OPP (C) with its slicing layers. Suppose OPP A is above over B, as shown in~\autoref{fig:OPPMerge}. 
Note that \textit{curving} cannot be applied for two OPPs with only closed \textit{contours}.

%Basic idea
Our basic idea is to produce a modified version of CurviSlicer to extract curved layers for the merged OPP C. There are two key questions: 1) how to determine the top/bottom \tfas of C? 2) how to guarantee the fabrication constraints? 
%For the slope angle constraint, we propose an assumption that as long as the upper and bottom layers meet the constraint, the in-between layers also meet the constraint. Based on that assumption, we propose to determine the \tfas and detect the slope angle constraint in one single step.
We solve the first question by combining the top/bottom \tfas of A and B. For the determined top/bottom \tfas of C, detect the slope angle constraint. Then extract the in-between curved layers with our modified CurviSlicer satisfying the fabrication constraints.

%step-by-step process
%1-slop constraint for the top (bottom) layers
For an OPP, define the projection of its top (bottom) layer as its top (bottom) \tfa (\revision{$A_{t}, A_{b}, B_{t}, B_{b}$}).
Starting traverse from the layer at the top layer of A, two oblique polylines can be obtained by connecting the two end points on both sides of each layer to the bottom layer. If both oblique lines violate the slope angle constraint, the two OPPs cannot be merged via curved layers. If not, we generate C's top \tfas by combining 1) OPP A's top \tfa (\revision{$A_{t}$}), 2) the difference result of B's top \tfa and A's bottom \tfa (\revision{$B_{t}-A_{b}$}), 3) the oblique lines 
that meet the slope constraint; C's bottom \tfas are taken from B's bottom \tfas.

%2-
%2-slope and thickness constraint for the in-between layers
Next, we call CurviSlicer, specifying both top and bottom \tfas.
\revision{CurviSlicer formulated two key terms in their objective: one flat term to determine whether the target area can be flattened with the slope angle and thickness constraints and one smooth term to make the generated curved layer smoother.
CurviSlicer with the smooth term would be much more time-consuming, which is unnecessary for our case, since the \textit{curving operation} would be called frequently, and the generated layers would not be used in the final toolpath generation. So we only use the flat term while applying CurviSlicer.}
%\revision{Different from} the original version of CurviSlicer, we only use the flat term to determine \revision{can be} flattened \revision{where the smooth term can be ignored} for the slope angle and thickness constraint detection. 
%We leave the final toolpath smoothing in a post-optimization process.
The merge is executable if the top and bottom \tfas are successfully flattened without violating the fabrication constraints.
\revision{Note that CurviSlicer only works for watertight 3D models. For the surface model, we convert it to an approximate watertight model with a minimal shell thickness that can be taken as the input of CurviSlicer (details in~\autoref{sec:tricks}).
}

\section{ACAP Methodology}
\label{sec:method}
This section describes our algorithm in more details. 
For clear exposition, we explain the methodology for open 2D patches with only \textit{segments} in each layer.
The extension to 3D will be discussed in \autoref{sec:extension}. 
\revision{As introduced in~\autoref{sec:overview}, the basic idea of our algorithm is a bottom-up OPP merging process based on a unified graph-based representation of the OPP graph and a set of graph nodes merging operations for the OPP graph. The OPP graph encodes the surface decomposition and their dependencies during the bottom-up OPP merging process. The OPP node merging operations are formulated through flat and curved slicing layers.}

\subsection{Building Dependency Graph}
\label{sec:dependency-graph}
\revision{With an orientation that meets the support structure constraint (see \autoref{sec:oridetail} for details), we uniformly slice the model with flat planers vertical to the printing direction by the layer thickness (1.0mm in ceramic printing, 0.2mm in FDM), then build a directed acyclic graph to describe the dependency relationships}, \revision{named \textit{dependency OPP graph} $G_{depend}$}, where each node represents a sliced element (\textit{segment}), \revision{each directed edge represents a dependency relationship between two neighboring nodes where the closest horizontal distance between them is smaller than the path width (6.0mm in ceramic printing, 1.5mm in FDM)}, shown in~\autoref{fig:InitialOPPGraph}(a).
If node $N_1$ has a directed edge pointing to $N_2$, \revision{it indicates that (1)} $N_2$ can \revision{only} be printed only after $N_1$ \revision{ and (2) the two OPP nodes can be merged through \textit{stacking} operation.} A node can be printed only if all nodes it depends on have been printed.

%In the orientation slicing process \hs{where is this process? define it before?}, we have obtained a set of continuous curves \hs{??} $L = \{l_{ij},i = 0,...,n, j = 0,...,m\}$ through the intersection of slicing plane and mesh, where $i$ represents the index of layers and $j$ stands for the index of the curves in $i$ layers. 
%Then we establish a dependency graph ($G_{depend}$), where every node represents a $l_{ij}$, and every directed edge represents the dependencies between nodes. For example, if node $A$ has a directed edge pointing to $B$, $B$ can be printed only after $A$, call it "$B$ depends on $A$". A node can be printed only if all nodes it depends on have already been printed.

%Note that we record the Euclidean distance between the terminal points of all the connected nodes, which will be used in Section~\ref{sec:PathGeneration}. For two nodes connected by edge, if they meet the precedence constraint, the $segments$ they represent can be printed by a continuous path even though their terminal points are far away. We will describe how to connect them in Section~\ref{sec:PathGeneration}.
%As we mentioned in Section 3.2, we divide all these curves into two different types: \emph{segment} and \emph{contour}. \emph{Segment} is a curve that has two terminal points while \emph{Contour} is a closed curve that have no terminal point. For a 2D mesh, there is only \emph{Segment} but not \emph{Contour}.Therefore, in this section, we describe only the processing of \emph{Segment}, the processing of \emph{Contour} will be described in Section 5.

\subsection{OPP Merging through Flat Layers}
\label{sec:I-OPP-Merging}

%\paragraph{I-OPP merging}
%In order to effectively reduce the computation, the next step is to merge the dependency graph nodes to gain an initial OPP graph. Then, we design a novel search algorithm for calculating PC-MPC. The optimal solution is usually not unique, every one corresponds to an OPP graph. \textit{Stacking} is utilized here.
%The definition of OPP at the flat layers level is further utilized here.

%goal: try to decompose with purely flat layers; that is decompose the dependecncy graph consider 1) dependency 2) decompose as few as possible;
%basic idea: solve a classical PC-MPC problem. 1） to make the problem trackable We first  simple the 2) we propose an dasdasd. The two key steps are detail in the next two paragraphs.

This section aims at maximal continuity provided by flat slicing, that is, merging the flat sliced elements of $G_{depend}$ to a minimal number of I-OPPs.
Each sliced element can be seen as an I-OPP, which can be merged by \textit{stacking} operation (\autoref{sec:oppdef}).
The merged I-OPPs should keep the dependency relationships formulated in $G_{depend}$.
Such merging process is indeed to find a path cover for $G_{depend}$ with the fewest paths considering the dependency relationships.

%To make the NP-hard problem tractable, 
We propose two key steps for the merging process: 1) merge the nodes of $G_{depend}$ that must appear on the same path in any minimum path cover in advance, then build an \textit{initial OPP graph} $G_{init}$ to reduce the size of $G_{depend}$. 2) merge the nodes of $G_{init}$ further by solving
\revision{the path cover problem with the dependency constraints}. 
Rather than running an approximation algorithm, we propose a searching-based method with a pruning strategy to explore the possible solutions.  

%Having built $G_{depend}$, we can merge into a minimum number of OPPs composed of flat layers by solving PC-MPC problem. However, it is time-consuming to solve the NP-hard problem when $G_{depend}$ contains many nodes. To reduce the size of the graph, we merge the nodes that must appear on the same path in advance. \llu{??} 
%The new graph is called initial OPP graph ($G_{init}$). Then we further merge nodes in $G_{init}$ by solving a PC-MPC problem. Each path of path cover represents an OPP composed of flat layers (I-OPP).

\begin{figure}[t]
\centering
\includegraphics[width=0.9\linewidth]{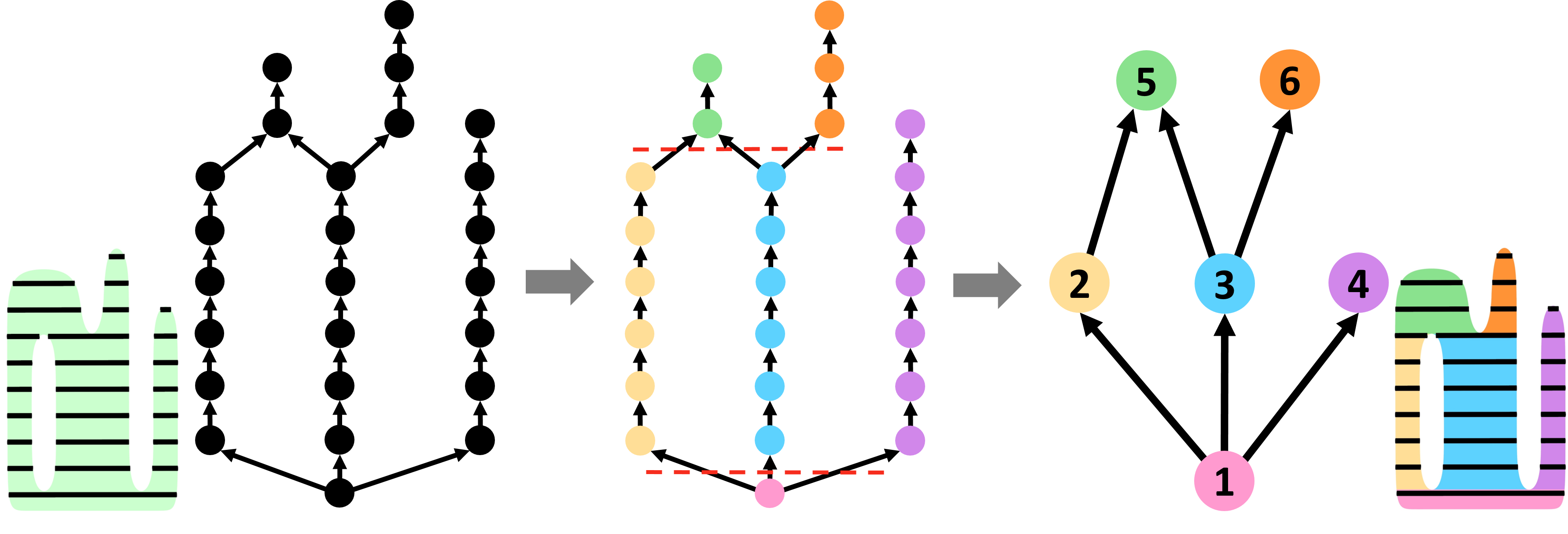}
%\vspace{-8pt}
\leftline{ \hspace{0.18\linewidth}
            \footnotesize{(a)} \hspace{0.235\linewidth}
            (b)\hspace{0.27\linewidth}
            (c)}
\caption{\revision{(a): \textit{Dependency graph} ($G_{depend}$), each node represents a sliced element, and each edge represents a dependency relationship between two nodes. (b): Temporarily ignore edges pointing to (from) nodes with in-degree (out-degree) not less than 2
(crossed by dashed red lines), and compute the connected components. (c): \textit{Initial OPP graph} ($G_{init}$), each connected component of (b) acts as a node and edges are inherited from $G_{depend}$.}}
\label{fig:InitialOPPGraph}
\end{figure}

\paragraph{Initial OPP Graph}
 To build a simplified graph $G_{init}$ from $G_{depend}$, we traverse all nodes of $G_{depend}$, if there are two or more edges pointing to the same node or starting from the same node, delete these edges temporarily (crossed by dashed red lines \revision{in~\autoref{fig:InitialOPPGraph}(b)}). 
 Then compute the connected components. Each component acts as a node of $G_{init}$, which can be seen as a merged larger I-OPP by \textit{stacking}.
 Such a merging process maintains the optimality, i.e., the sub-nodes of a $G_{init}$ node must belong to a single path of the optimal solution\footnote{This can be proved \revision{using proof by contradiction}. If two adjacent sub-nodes belong to \revision{the} two paths of an optimal solution, they must be terminal nodes of the paths. Obviously, the two paths can be \revision{further} connected, which \revision{shows that} the current path cover solution is not optimal.
%Therefore, when they are in the same path, the number of paths of path cover must be less than when they are not in the same path.
}.
 The dependencies of $G_{init}$ are inherited from $G_{depend}$. As shown in \revision{\autoref{fig:InitialOPPGraph}(c)}, the \revision{nodes are} significantly reduced. 
  %must be located on the same path in any minimum path cover, 

 %Figure~\ref{fig:PathCoverSolutionSpace}(b) shows corresponding $G_{init}$ from Figure~\ref{fig:PathCoverSolutionSpace}(a).
%The next step is to establish an initial OPP graph ($G_{init}$).
%The $G_{init}$ obtained is shown in Figure \ref{fig:PathCoverSolutionSpace}(b).
%the dependency between the nodes will be preserved and will still be represented by directed edges.

%We further merge nodes in $G_{init}$ based on the flat OPP definition. A sufficient condition for a patch to be a flat OPP is to conform to monotonicity. For any printable(meet precedence constraints) path $P$ in $G_{init}$, merging all the nodes contained in it, we can get an OPP that satisfies monotonicity. Since our goal is to partition the least OPPs, in there is equivalent to the problem of minimum path cover with precedence constraints, that is, covering all nodes in $G_{init}$ with the least number of paths with precedence constraints. 
    %As we said before, flat OPP needs to satisfy monotonicity.
    %If a node's in-degree or out-degree is greater than 1, all its exit (or enter) edges are deleted.
    
\begin{figure}[t]
\centering
\includegraphics[width=0.95\linewidth]{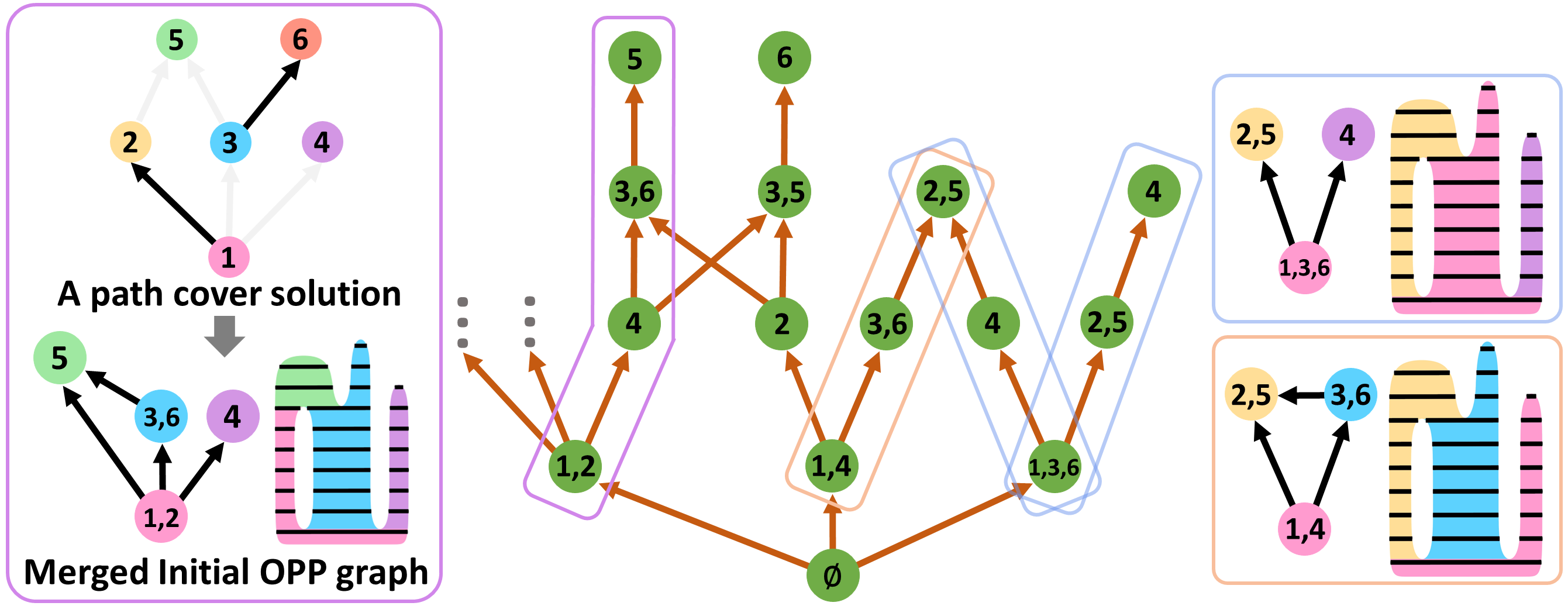}
%\vspace{-8pt}
\leftline{ \hspace{0.133\linewidth}
            \footnotesize{(a)} \hspace{0.32\linewidth}
            (b)\hspace{0.292\linewidth}
            (c)}
\caption{\revision{(a): Based on a path cover solution, the nodes of $G_{init}$ are merged. 
(b): 
The path cover solution space ($G_{solution}$), where nodes (green dots) encode paths (merged I-OPPs) of path cover solutions,
directed edges encode the printing order of the merged I-OPPs,
and each path cover solution corresponds to a sequence of directed edges and nodes. 
Different sequences of $G_{solution}$ may produce the same path cover solution but in a different printing order (two blue boxes).
(c): Three sequences (two blue and one orange boxes) produce two optimal path cover solutions associated with the minimal number of $G_{flat}$.}}
\label{fig:PathCoverSolutionSpace}
\end{figure}
    
\paragraph{\revision{Path Cover of Initial OPP Graph}} 
%As described before, PC-MPC problem can be converted into a PC-TSP problem in linear time, and PC-TSP problem is an NP-hard problem. 
%From $G_{depend}$ to $G_{init}$, the scale of the graph is significantly reduced. 
%Rather than running an approximation algorithm, we developed a novel search algorithm (Algorithm~\ref{alg:three}) with pruning skills to compute all minimum path cover in $G_{init}$. 
%Each node of $G_{OPP}$ is comprised of a sequence of I/II-OPPs (denoted as \textit{sub-OPPs}), where the sequence edges indicate the printing order. 
%basic idea of this algorithm:solution space formulation
%describe the general process of explore the solution space
OPP nodes of $G_{init}$ can be further merged via \textit{stacking}, \revision{as (1,2), (1,3,6), (3,5) in \autoref{fig:InitialOPPGraph}(c)}. 
\revision{
A merged \textit{flat OPP graph} $G_{flat}$ of $G_{init}$ can be generated as a result of a path cover solution of $G_{init}$, where each path proposes a merged I-OPP. 
Each node of merged OPP comprises a sequence of I/II-OPPs (denoted as \textit{sub-OPPs}), where the sequence edges indicate the printing order.
\autoref{fig:PathCoverSolutionSpace}(a) shows a merged $G_{flat}$ with four merged I-OPPs based on the  path cover solution \{(1,2), (4), (3,6), (5)\}.
Note that the dependency relationships of $G_{init}$ are preserved.} Such path cover solution can be generated from a specific Depth First Search (DFS) starting from the \revision{root nodes} of $G_{init}$. 
Different from a general DFS, it may not explore as deep as possible to the \revision{unexplored} node before backtracking. The search has to stop while a node's dependent nodes have not been explored, such as (1,2,5) is not valid that (3) is not explored. 

%is built along with the exploration procedure
%\revision{As shown in \autoref{fig:PathCoverSolutionSpace}},
\revision{The path cover solution space of $G_{init}$  can be represented by a specific directed graph structure ($G_{solution}$)}, where the root node is set to an 
\revision{empty node (an additional virtual node)}, the non-root nodes indicate the corresponding merged I-OPPs of path cover solutions, the directed edges encode the printing order of merged I-OPPs.
\revision{
A path cover solution is presented as a finite sequence of directed edges and nodes of $G_{solution}$ starting from its root node, (see the sequence of $\{(1,2), (4), (3,6), (5)\}$ in \autoref{fig:PathCoverSolutionSpace}(b)).
}

\paragraph{\revision{Path Cover Solution Space Exploration}} 
%propose two pruning operations
%1: choose the greed strategy ;
%2: branch and bound technique to pruning
\revision{To merge $G_{init}$ into the minimal number of I-OPPs, we aim to search for the shortest sequences while exploring $G_{solution}$.}
%A straightforward but time-consuming method is to traverse $G_{solution}$ by attempting all possible branches.
We propose \revision{two} key techniques to speed up such exploration by pruning the solution space. 
\revision{First, starting from the root node of $G_{solution}$, we apply a beam search procedure with the branch and bound technique, to explore $G_{solution}$ level by level, where the beam search width is set to $W = 10^4$ in our implementation.}
\revision{For each level of $G_{solution}$, we sort its candidate nodes by the number of included $G_{init}$'s nodes, which implies that we tend to pick the nodes of $G_{solution}$ including $G_{init}$'s nodes as many as possible.
Note that nodes of $G_{solution}$ would point to the same node of next level if its sequence included $G_{init}$'s nodes remain the same, such as \{(1,2), (4)\} and \{(1,4), (2)\} both point to (3,6) and (3,5).}
\revision{Second, for generating candidate nodes of each beam search iteration, we use a greedy strategy in the path cover solution generation with DFS, that each traversal would explore as deep as possible, as the traversal (1,3) will not terminate at node (3) since node (6) can be added into (1,3,6) shown in \autoref{fig:InitialOPPGraph}(c). We have proved the optimality of this strategy}\footnote{In the traversal path process, if terminating the path when a node can be added, the node must be the starting point of another path. In the same way as the last proof, the number of paths of path cover, in this case, is at least 1 more than our strategy, so it is not optimal.}.
\revision{The proposed path cover exploration method would produce multiple optimal $G_{flat}$ with the least number of OPP nodes shown in  \autoref{fig:PathCoverSolutionSpace}(c)}.
The pseudo-code is presented in Appendix~\ref{sec:pseudocode}.

\subsection{OPP Merging through Curved Layers}
\label{sec:combination}

\begin{figure}[tb]   
\centering
\includegraphics[width=0.8\linewidth]{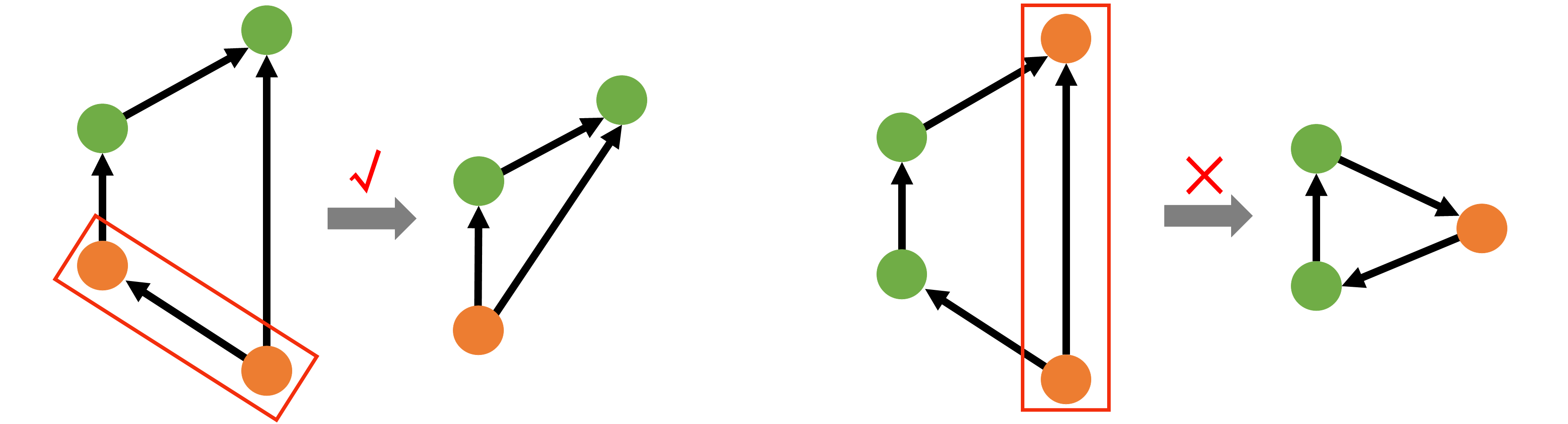}
%\vspace{-8pt}
\caption{Illustration of printing dependency deadlock, the orange nodes represent the pair of selected nodes in DAG.
\revision{Left: Only one path exists between the two nodes, and no deadlock occurs after merging. Right: Two or more paths between the two nodes may produce deadlock loops after merging, i.e., the nodes are inter-dependent; thus, merging is prohibited.
}
}
% Strategy for merging a pair of nodes (orange ones here) in $G_{OPP}$. \hs{no, this figure is to illustrate the dead-lock. no?}
% Left: Only one path exists between the two nodes, allowing merging.   
% Right: More than one path exists between the two nodes, resulting in deadlock, thus prohibiting merging.
%\hs{caption needs updating with more contents}}
\label{fig:deadlock}
\end{figure}

\begin{figure}[t]   
\centering
\includegraphics[width=1.0\linewidth]{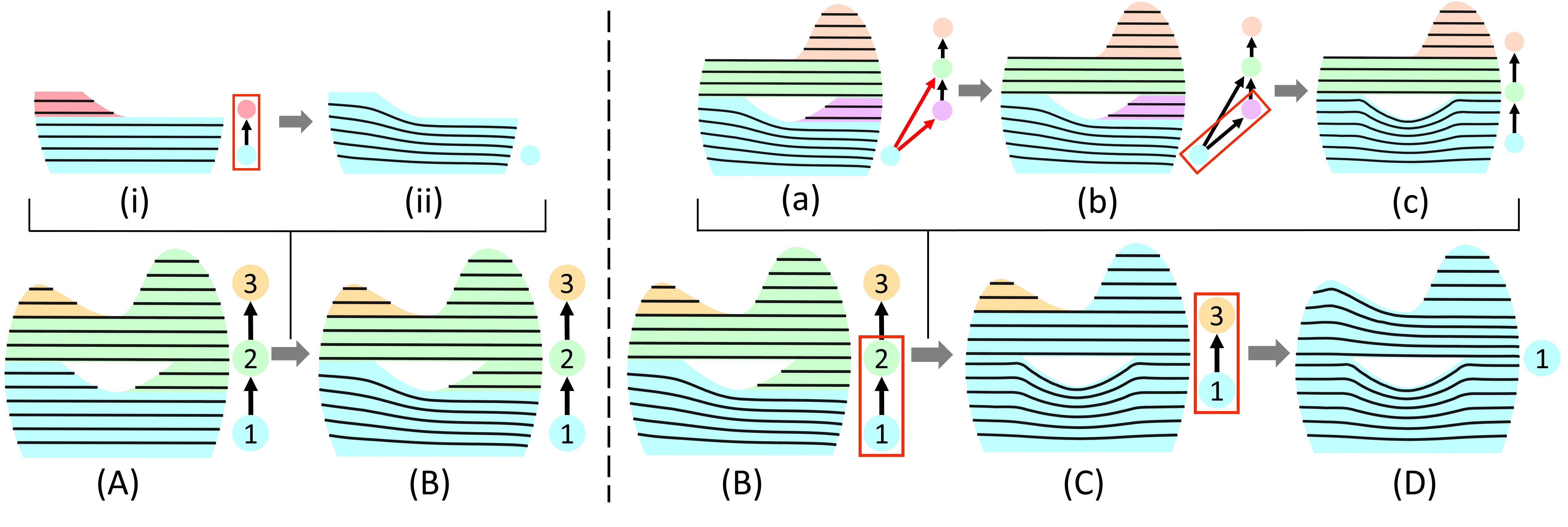}
%\vspace{-8pt}
\caption{A mesh is decomposed into three I-OPPs, corresponding to \revision{$G_{curved}$} with three nodes (\revision{A}).
\revision{First, in the initial merging process, we merge the sub-OPPs within each OPP by \textit{curving} operation, indicated in (A, B), where (i, ii) show the merging process of sub-OPPs within OPP 1. %Note that the sub-OPPs of OPP 2 and 3 could not be merged.
Then, in the OPP merging process, we iteratively merge nodes pair by pair to one III-OPP node (B$\thicksim$D).} 
\revision{a$\thicksim$c} show the merging process in the inner loop of (\revision{B}) to (\revision{C}), and only the three nodes connected by red edges in (\revision{a}) are selected to merge. Since the \revision{three} sub-OPPs of (\revision{c}) accept \textit{stacking} into one OPP, node \revision{1} and \revision{2} can be successfully merged.}
\label{fig:iterative merge}
\end{figure}

% \begin{figure}[t]   
% \centering
% \includegraphics[width=1.0\linewidth]{images/violate rules.png}
% \caption{Since the purple patch and the green patch meet \textit{curving} rules (left), they can merge into a II-OPP (right), but it's top layer projection on the mesh does not intersect with the projection of the bottom layer of yellow patch. The yellow and green patches are unable to merge into a III-OPP (right).}
% \label{fig:violate rules}
% \end{figure}

%It is worth mentioning that we apply the \textit{curving} operation as much as possible \revision{considering curved layer helps reduce the staircase defect (both in \textbf{initial merge} and \textbf{inner loop}), as mentioned in \autoref{sec:oppdef}.
%For example, as shown in \revision{Figure~\ref{fig:iterative merge}(b)}, although these OPPs can be merged via \textit{stacking}, we perform \textit{curving} to \revision{(c)}.}

Up to now, we obtained a set of unique $G_{flat}$ \revision{with the maximal continuity via flat slicing. Could we further merge its OPP nodes? Recall that we observe curved slicing layers can be exceptionally effective in generating continuous printing paths for multiple separate printable surface patches of flat layers. Driven by this insight, we apply the \textit{curving} operation (defined in~\autoref{sec:judgement}) \textit{as much as possible} to further reduce number of OPPs. This can be done in two steps. First, apply the \textit{curving} operation to sub-OPPs of each OPP node (\textbf{Initial Merging Process}), which would enlarge the target flat areas to be beneficial to the subsequent OPPs merging process. Second, apply the \textit{curving} operation to merge multiple OPPs (\textbf{OPP Merging Process}).
Below we propose a general pairwise-based merging procedure for both initial and OPP merging processes.
We set $G_{flat}$ as an initial \textit{curved OPP graph} $G_{curved}$, iteratively merge $G_{curved}$ to implement the two merging processes.
II-OPPs and III-OPPs would be generated accordingly.}

\paragraph{\revision{General Pairwise Merging Procedure}}
\revision{OPP graphs are directed acyclic graphs (DAG). We propose an iteratively pair-wise merging procedure for a general DAG.}
For each iteration, randomly select an edge and try to merge the two related nodes with specific merging criteria to check whether they can be merged or not. If yes, update the graph by \revision{1) erasing the edges of the two nodes, 2) merging a pair of nodes to one node} and 3) \revision{connecting} other related edges of the two nodes to the merged node.
The terminal condition is that no pairs of nodes can be merged. 

\paragraph{\revision{Merging Criteria of OPP Graph}}
In our case, the directed edges of OPP graph indicate dependency relationships of OPPs. While merging a pair of nodes, one necessary criterion is to guarantee there are no multiple paths between the two nodes, since it will result in a deadlock of dependency after merging, shown in~\autoref{fig:deadlock}(right). 
\revision{Another criterion is that the \textit{curving} operation can be applied to the two OPPs (sub-OPP) nodes.
The two merging processes are demonstrated in~\autoref{fig:iterative merge}(A$\thicksim$B) and (B$\thicksim$D). For each iteration of \textbf{OPP Merging Process}, we aim to merge two OPP nodes with the \textit{curving} operation, which is taken as an inner loop of the \textbf{OPP Merging Process} (see a$\thicksim$c in~\autoref{fig:iterative merge}).} 

%For the \textbf{Initial Merging Process} (\autoref{fig:iterative merge}(A$\thicksim$B)) and \textbf{outer loop} (\autoref{fig:iterative merge}(B$\thicksim$D)), since they maintain a DAG, the pairwise merging strategy can be applied directly.}

%mentioned apply the strategy in two levels.
\paragraph{\revision{Inner Loop of OPP Merging Process}}
As for the inner loop, a DAG can be formulated from the two candidate merging OPP nodes: 1) take their sequences of sub-OPPs where each sub-OPP is a node and maintain the dependency edges of these sub-OPPs; 2) add back the associated dependency edges between the two sequences from $G_{init}$.
While applying the pairwise merging strategy to the result DAG \revision{(\autoref{fig:iterative merge}(a$\thicksim$c))},
we only select edges that benefit merging the two sequences to a single sequence. Specifically, we label the nodes that have edges across the two sub-OPP sequences (two red edges shown in \revision{(a)}), then select edges over such labeled nodes.
%as the two red edges shown in Figure~\ref{fig:iterative merge}(d).
For each pair of nodes, call \textit{curving} operation to merge the two nodes (\autoref{sec:judgement}).
The pseudo-code is presented in Appendix~\ref{sec:pseudocode}.

%Similarly, each time a pair of $G_{OPP}$'s nodes is selected to merge, we take out their sub-OPPs and add back the associated edges between them from $G_{init}$, using the same merging strategy for the graph thus constructed. 
%We essentially apply this merging strategy on two nested levels, $G_{OPP}$ is located in outer loop level (shown in Figure ~\ref{fig:iterative merge}(a) $\thicksim$ (c)), and the graph composed of sub-OPP is in inter loop level which nested in outer loop level.
%(shown in Figure ~\ref{fig:iterative merge}(d) $\thicksim$ (g)), the merge of outer loop level's nodes is based on merging inter loop level's nodes.

%Now we describe the key differences for two levels, respectively. 

%difference of outer loop level: merge
%In outer loop level,  when trying to merge two selected nodes, we looking at whether the remaining inter loop level's nodes can \textit{stacking} into one OPP after it's iterative process terminates, if yes, then merge. 

%difference of inter loop level: select nodes, merge

%discussion: different G_opp may own different number of OPPs; we curving as much as possible, while still exist different combination of II-OPP; different order can also result in different final results.

\begin{wrapfigure}[9]{r}{0.135\textwidth}
\vspace{-15pt}
\hspace{-25pt}
\centering
\includegraphics[width=0.15\textwidth]{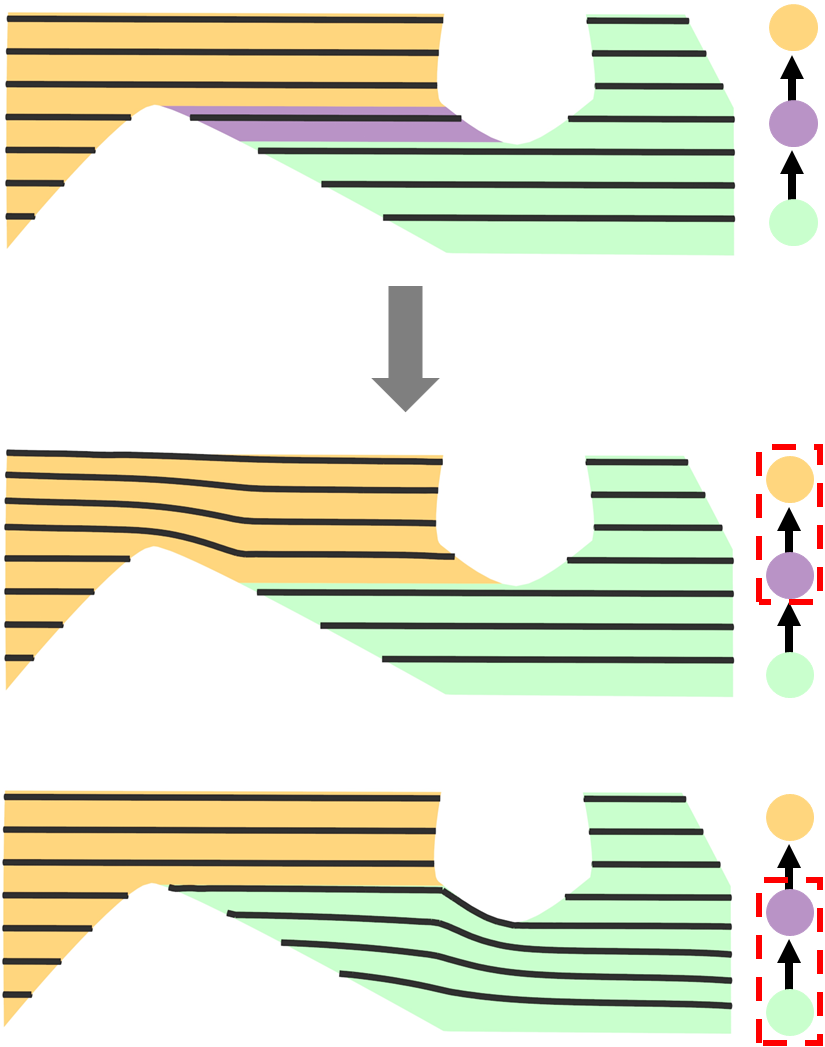}
\end{wrapfigure}

\paragraph{\revision{Optimality of Proposed Method}}
For different \revision{$G_{curved}$} of input, the final number of nodes after merging may be different. We randomly select one \revision{$G_{curved}$} with the least number of nodes since we only consider the OPP number criterion. Different \revision{$G_{curved}$} and order of merging OPPs by \textit{curving} may result in curved layers with different distributions, in other words, different sub-OPP of the final III-OPP, as shown in the inset.
Similarly, since the order of selecting nodes in two levels is random, we cannot guarantee the global optimal solution. \autoref{fig:failure case 1} shows an example, and more details are discussed in \autoref{sec:Different merging results}.

\subsection{Layers Connection}
\label{sec:PathGeneration}
With the bottom-up merging OPP process, we get an optimal OPP decomposition. This section works on path planning for each OPP by converting OPP slicing layers to continuous toolpath and generating transfer moves between OPPs.
Note that we remove the most time-consuming smooth term of CurviSilcer while applying the \textit{curving} operation during OPP merging (\autoref{sec:judgement}). Here we add it back and rebuild the smoother curved layers of related OPPs for the toolpath planning.
Then connect inter-layer to a single path for each OPP and determine the fabrication order of these OPP paths.

\paragraph{Inter-layer Connection Path}
Since there are only $segments$ in 2D models, they can be connected using Zig-zag pattern. 
\revision{For the two terminal points of a \textit{segment}, if one is the entry point, the other will be the exit point. We select entry points for all $segments$, minimize the total length of the connection path between the entry and exit points of adjacent $segments$.}
Then we set a Euclidean distance threshold $D$ to determine whether two terminal points of two neighboring layers can be connected directly with a straight segment. 
\begin{wrapfigure}[4]{r}{0.13\textwidth}
\vspace{-10pt}
\hspace{-20pt}
\centering
\includegraphics[width=0.165\textwidth]{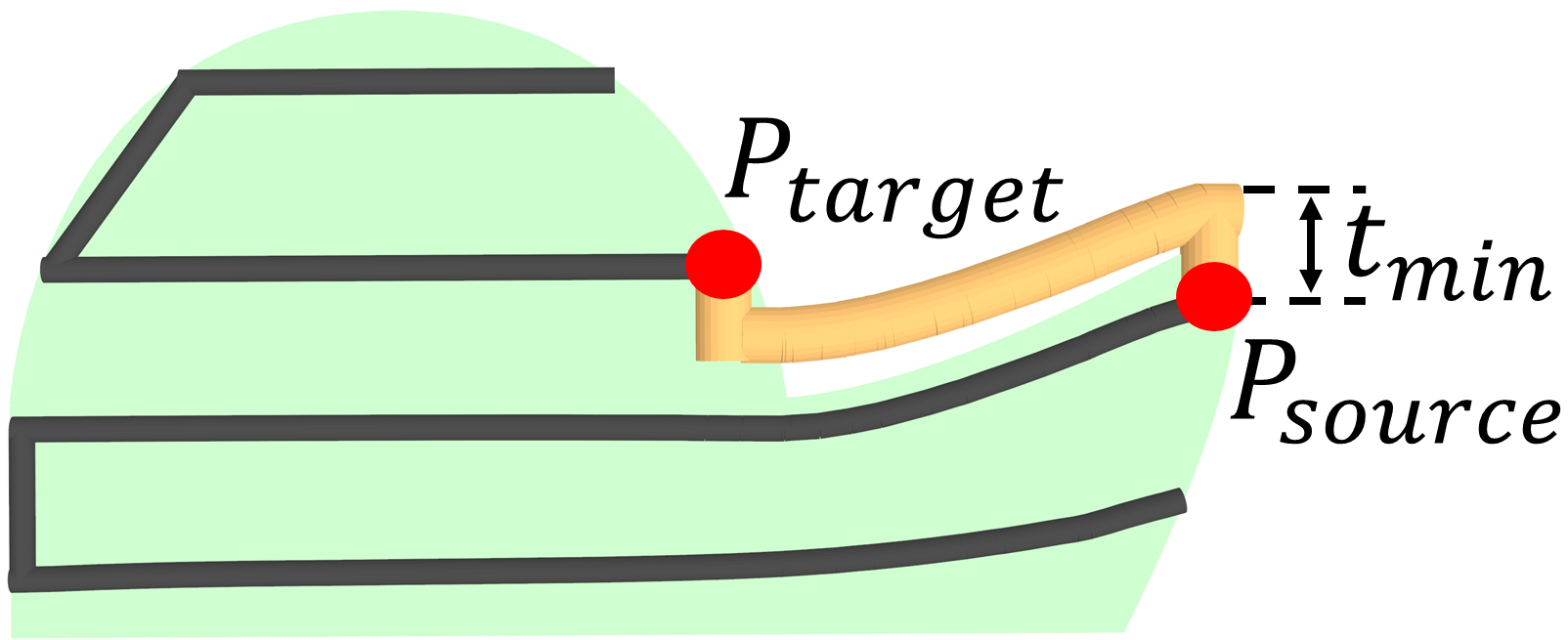}
\end{wrapfigure}
If the distance exceeds $D$, an extra path is added. See the inset, call the current terminal point in the current layer $P_{source}$, and the terminal point to be connected in the next layer $P_{target}$. Print along the current printed layer with $t_{min}$ layer thickness from $P_{source}$ to the position closest to $P_{target}$, and then print along straight line to $P_{target}$ (the orange path in the inset). 
The post path optimization in \autoref{sec:extension} will improve the spatial distribution of the generated extra path.
% \revision{
% Still, we remark that this extra path might be outside the model, and such an artifact cannot be guaranteed to be avoided.}
% The Inter-layer connecting path can be outside of the model by a distance of t_min, while it is hypothetically possible to reroute the paths below it to circumvent such a defect. This should be mentioned as a limitation of the current approach.

\paragraph{OPPs Sequence} 

%%%Because the machine needs to stop and start urgently at the turning point, there will be a very short pause, resulting in additional extrusion of clay. The extrusion amount can be reduced through the program setting to offset this effect.
%The last step we compute the manufacturing sequence of OPPs.
With the known dependency relationships of these OPP paths, we apply a method as~\autoref{sec:I-OPP-Merging} to produce a feasible fabrication order. 
Then, we plan travel moves between OPP paths by withdrawing the nozzle to a safe distance above the printed objects to avoid possible collisions (see~\autoref{fig:teaser}). Finally, a G-code file is generated to transfer the toolpath to the printer.

%There are also dependency constraints defined by $G_{depend}$ between OPPs. This is a classic PC-TSP problem and we use greedy method to solve it approximately.
%In addition, from one OPP to another, we plan travel move of the nozzle. We simplify the collision problem caused by the nozzle in this process, and adopt a slender nozzle to avoid collision. 
%\cite{li2021multi} provide a more rigorous solution about it.
%However, we do not consider the collision problem caused by the nozzle in this process and adopt a slender nozzle to alleviate collision. \cite{li2021multi} provide a better solution about it.
%The coordinates of sampling points along the path and extra machine information (nozzle on/off, pump speed, and feed value) are written into a G-code file, which can be used for printing directly.

{\color{black}
\section{Global Collision Considerations}
\label{sec:extendToGeneralNozzle}
The algorithm of~\autoref{sec:method} does not consider the global collision raised by the printed parts exceeding the nozzle length shown in~\autoref{fig:collision}(a), which often occurs when printing a lower layer after the higher parts.
%basic idea
To extend our method with this consideration, the key challenge is how to represent the global collision constraint in our proposed bottom-up OPP merging algorithm. 
We intend to formulate such constraints to a new type of directed edges of the OPP graph, named "collision dependency edges."  Similar to the original directed edges, the new edges would represent printing dependencies. The difference is that the OPP nodes of the new edges cannot be merged through \textit{stacking} operations. We would first clarify the generation of collision dependency edges and then introduce the algorithm's modifications raised by the new edges.

\begin{figure}[tb]   
\centering
\includegraphics[width=1.0\linewidth]{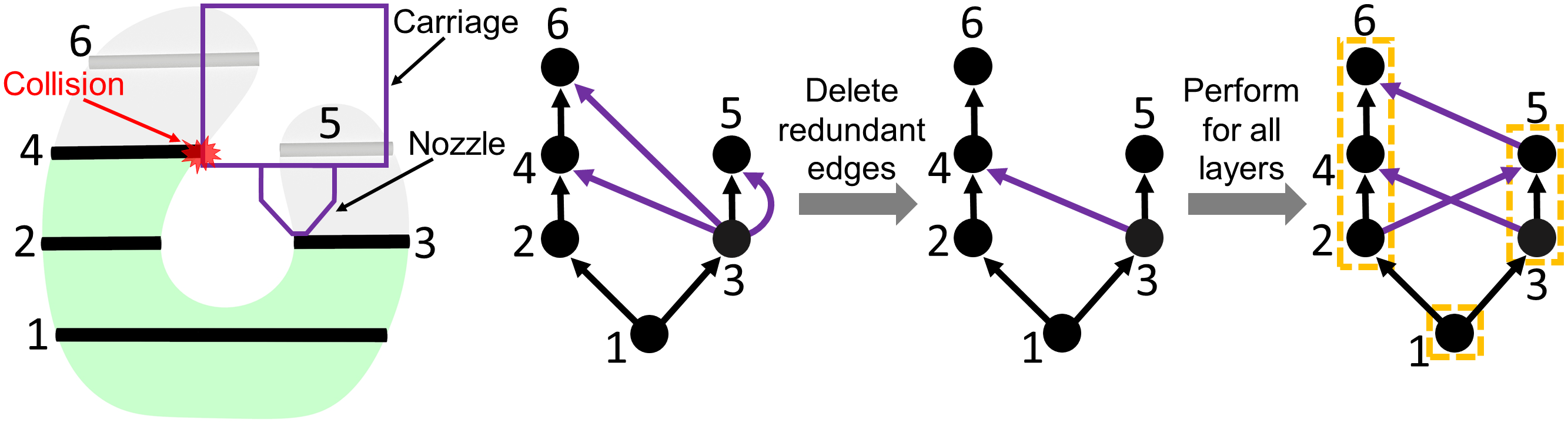}
%\vspace{-8pt}
\leftline{ \footnotesize  \hspace{0.11\linewidth}
            {(a)} \hspace{0.225\linewidth}
            (b)\hspace{0.225\linewidth}
            (c)\hspace{0.215\linewidth}
            (d)}
\caption{\revision{Illustration of collision detection and building \textit{dependency graph}. For layer "3", traverse and perform collision detection with other layers, and collision occurs (a). (b): The layer "3" will collide with three layers, thus three candidate collision dependency edges (purple) are supposed to insert in  $G_{depend}$. (c): Delete two redundant candidate collision dependency edges. (d): Perform the method for all layers and finally three collision dependency edges are added to $G_{depend}$. The orange boxes show the merging results in $G_{init}$ without adding collision dependency edges.}}
\label{fig:collision}
\end{figure}

\paragraph{Generation of Collision Dependency Edges}
To append the novel edges of $G_{depend}$, 
we apply the same method to model the printing nozzle in~\autoref{sec:fabcon}. %The nozzle's carriage is taken as a 3D box.
%and the devices above the carriage is an infinite platform.
For each node pair of $G_{depend}$, add a collision dependency edge between them if a collision occurs while printing. As the layer "3" and "4" in~\autoref{fig:collision}(a), the collision indicates that the layer "4" must be printed after the layer "3". 
Note that we maintain $G_{depend}$ as a Hasse diagram~\cite{HasseDiagram}, where redundant dependency edges do not exist. For example, if exist two edges: A$\rightarrow$B and B$\rightarrow$C, then A$\rightarrow$C is a redundant edge.
%If a candidate collision dependency edge is redundant for $G_{depend}$, it will be ignored. 
The example in~\autoref{fig:collision}(b) shows that three candidate collision dependency edges starting from the node "3" has been compressed into one single edge (c).
%(d) shows the resulting $G_{depend}$.

\paragraph{Algorithm Modifications}
1) For~\autoref{sec:judgement}, we need to add a requirement for applying \textit{curving} operation, to guarantee no collision dependency between the top and bottom target flat areas of two OPPs.
2) For~\autoref{sec:I-OPP-Merging}, the method of path cover solution exploration remains the same as above. However, suppose a node of path cover solution space ($G_{solution}$) includes sub-OPP nodes with "collision dependency edges", these sub-OPP nodes cannot be merged in $G_{flat}$. 
3) For~\autoref{sec:combination}, we observe that the number of OPP nodes (6 nodes) in $G_{init}$ becomes much more than that without including collision dependency edges (3 nodes indicated with orange boxes), as shown in~\autoref{fig:collision}(d). 
To solve the efficiency problem raised by the increasing of OPP nodes, especially for the OPP merging procedure with curved layers, we add an external step before the \textit{Initial Merging Process}. That is to apply the \textit{stacking} operation for sub-OPPs of each OPP node according to the $G_{init}$ which is generated without considering the global collisions. 
}

\section{Extension to 3D}
\label{sec:extension}
Extending our algorithm from 2D to 3D does not require extra effort for most steps, except to deal with \emph{contour} in the $G_{init}$ construction step and extend the toolpath generation step for the 3D case. In this section, we describe the extension of these steps in detail.
% \revision{Note that the input 3D surface model is mainly used in three steps: 1)~\autoref{sec:dependency-graph} where we slice the input model; 2)~\autoref{sec:judgement} where we generate curved layers through CurviSlicer; 
% 3) “Toolpath optimization” in this section where we map the toolpath back to the input surface during optimization.}\llu{for what?}

\paragraph{Initial OPP Graph}
For 3D cases with \textit{contours}, we first build a $G_{init}$ with the method of~\autoref{sec:I-OPP-Merging}. If an OPP node of $G_{init}$ has both \textit{segments} and \textit{contours}, divide it into pure \textit{segment} nodes and pure \textit{contour} nodes. Such classification is conducive to the next step of merging via \textit{curving}, which only allows the input of two OPPs composed of \textit{segments}, or one OPP composed entirely of \textit{segments}, and the other entirely of \textit{contours}.

\paragraph{Contour Spiralization}

\begin{figure}[t]
\centering
\includegraphics[width=1.0\linewidth]{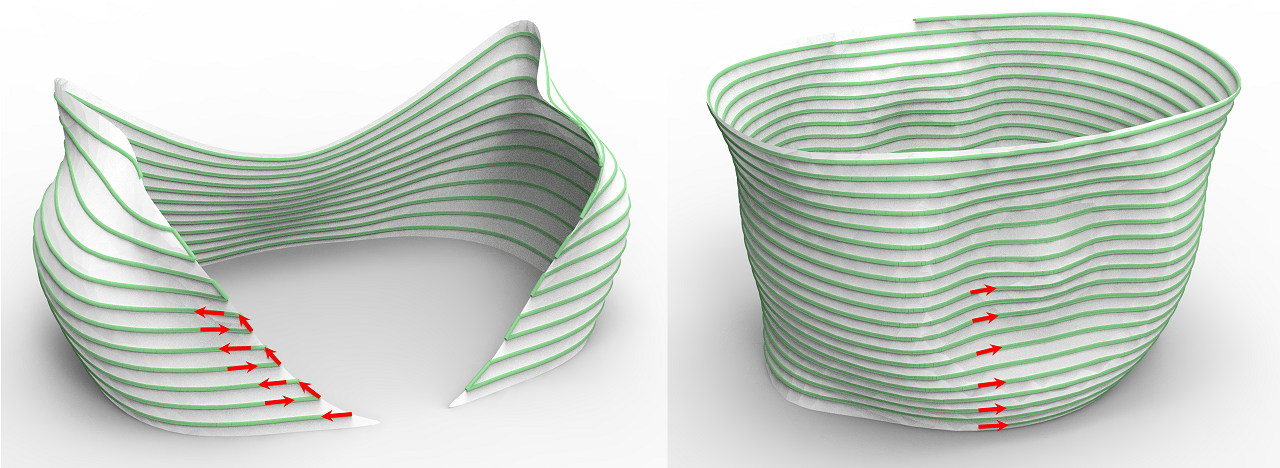}
%\vspace{-8pt}
\caption{Zig-zag connecting path for $segments$ (left) and Spiral connecting path for $contours$ (right).}
\label{fig:zigzag and spiral}
\end{figure}

%%%%%Fermat spiral is a strictly continuous path, at the same time, it can directly connect to other flat or curved path, but in a three-axis printer, it also has to meet the slope angle constraint. (describe in intro)
%%%%%For "contour nodes" merged by judgement algorithm, we extracted the patch they represent from original mesh. Since the slope constraint is conformed when printing in all directions on this patch, Fermat spiral \cite{Zhao2016}path is available. We generate it in the 3D patch.
%For "contour nodes" merged by judgment algorithm, since the slope constraint is conformed when printing in all directions on this patch,  we use connected Fermat spiral \cite{Zhao2016} to generate a continuous closed path.

%As mentioned, $segments$ of an OPP would be connected following a Zig-zag pattern.
%we connect them by Spiral path, shown in~\autoref{fig:zigzag and spiral}.
%Figure~\ref{fig:zigzag and spiral} (left) and (right) show the Zig-zag and Spiral path over 3D surface, respectively.

%To formulate the objective function for curve optimization, we discretize the input curve via curvature- based sampling so that we place more samples around high- curvature sections.

\revision{
%Every \textit{contour} is composed of $m$ discrete points, these points are all optional terminal points to choose.
To spiralize \textit{contours} ($C_{1}$,...$C_{n}$), we need first to determine one connecting point for each \textit{contour}. As~\autoref{sec:PathGeneration}, we aim to minimize the total distance between connecting points of adjacent \textit{contours}. 
We design a dynamic programming algorithm to choose appropriate connecting points for each \textit{contour}.
We discretize each \textit{contour} with $m$ sampling points, where $P_{ij}$ is a sampling point in $C_{i}$.
Denote $d_{ij}$ as the minimum sum of length between neighboring \textit{contours} $C_{1}$,...$C_{i}$ when choosing $P_{ij}$ as connecting point. The transition equation is as below.
    \begin{equation*}
        d_{ij}=\left\{\begin{array}{l}
        \ \ \ 0, \ i=1 \\
        \mathop{min}\limits_{k=1 \cdots m}(d_{(i-1)k}+\left\|P_{(i-1) k},P_{i j}\right\|_{2}), \ { i\neq 1} 
        \end{array}\right.
    \end{equation*}
Then, we connect these \textit{contours} by spiral path, shown in~\autoref{fig:zigzag and spiral}. For each two adjacent \textit{contours}, interpolate all sampling points starting from the connecting points:} 
% \hs{After determining the connection points of all contours, we connect them by spiral path, shown in~\autoref{fig:zigzag and spiral}.}
% The basic idea is to extract an interpolation curve between each two adjacent contours, then connect these curves to a single spiral. %where the method of curved layers is different from flat layers.
% First, determine an entry point for each contour where the key consideration is to minimize the total distance between adjacent entry points. 
% Second, for each two adjacent contours, interpolate all sampling points starting from the entry points: 
\begin{equation*}
  P_i = w \cdot P_a + (1-w) \cdot  P_b,\quad w = S_{current}/S_{total}
\end{equation*}
where $P_b$ is a sampling point of the below \textit{contour} \revision{(if $S_{current}=0$, $P_b$ is connecting point of current \textit{contour})}, $P_a$ is the nearest point to $P_b$ of  the above \textit{contour}, $S_{total}$ is total length of the below \textit{contour}, $S_{current}$ is the geodesic distance that $P_b$ moves from the connecting point along the below \textit{contour}, $P_i$ is the interpolation point. \revision{Note that the top \textit{contour} as a boundary will not be spiralized.}

\paragraph{Filling the Low-Slope Area}

\begin{figure}[t]
\centering
\includegraphics[width=1.0\linewidth]{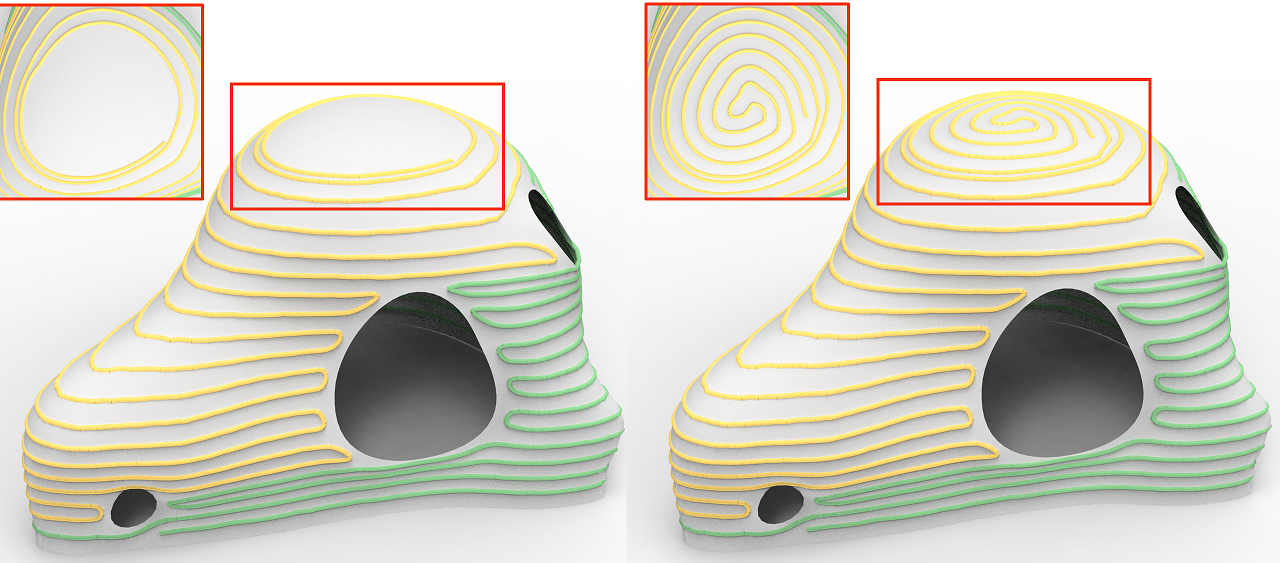}
\caption{Under-fills appears at the top of the model because of inadequate slicing in low-slope region (left). Filling the area with connected Fermat spirals (right), the top of model is printed completely.}
\label{fig:FermatSpiral for filling}
\end{figure}

\begin{figure}[t]   
\centering
\includegraphics[width=1.0\linewidth]{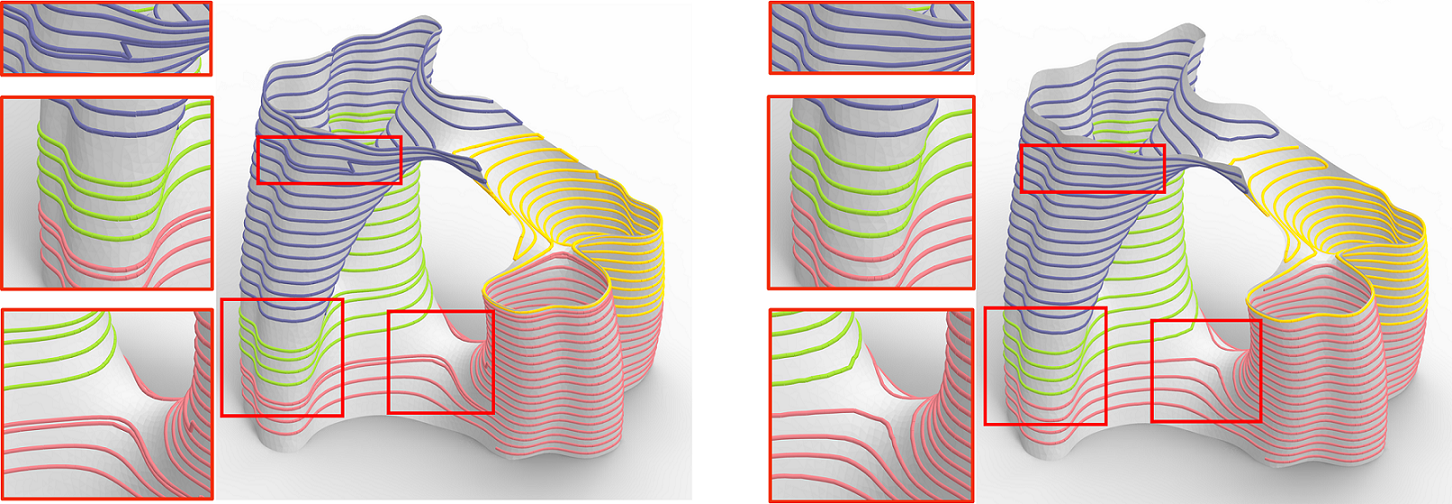}
\caption{Effectiveness of global path optimization. Left: without global path optimization, some paths are too close or too far. Right: global path optimization leads to uniform path. \revision{The toolpath optimization cannot guarantee a perfectly uniform solution as the area around the saddle point.}}
\label{fig:global optimization}
\end{figure}

While slicing, inadequate layers over low-slope regions inevitably result in under-fills, as shown in~\autoref{fig:FermatSpiral for filling}(left). Such under-fills of small areas can be removed by a post path optimization process (see \autoref{fig:global optimization} left bottom).
%As shown in~\autoref{fig:FermatSpiral for filling}(left), the layer thickness of clay based printing (the average is about 1.5$mm$) is much larger than that in plastic based printing (the average is about 0.2$mm$), which aggravates this problem.
%Global path optimization can handle this problem when the area is small, see Figure \ref{fig:global optimization} left bottom. 
For large under-fill areas, we propose to fill them with connected Fermat spirals~\cite{Zhao2016}. In our implementation, we set a threshold of $20\degree$ to detect the \revision{Fermat spiral region}. \revision{We traverse each pair of the adjacent \textit{contours}, 
generate the matching edges between the sampling points of two \textit{contours} by minimal Euclidean distance, and
measure the angle of these matching edges from the horizontal plane. If all angles of these matching edges are smaller than the threshold, add the surface patch between the two \textit{contours} to the Fermat spiral region.}
%We check all the adjacent flat layers, for each pair of nearest sampling points between them, calculate the slope angle of their connected straight line, if all exceeded the threshold of $20\degree$, the patches between them were judged as flatten area, merge the adjacent ones and generate 3D Fermat spiral path, which does not affect the number of OPPs. 
\autoref{fig:FermatSpiral for filling}(right) shows the filling path over the original 3D surface.

\paragraph{Toolpath Optimization} 
Since we generate toolpath for each OPP separately, the adjacent paths may be too close or too far from each other (left center of \autoref{fig:global optimization}(left)). 
Some inter-layer connection paths (\autoref{sec:PathGeneration}) and the filling paths of low slope areas (left bottom of \autoref{fig:global optimization}(left)) may be not uniformly distributed (left top of \autoref{fig:global optimization}(left)). 
%In addition, the filling paths on low slope area may cause obvious under-fill area because of insufficient sampling rate (left bottom of Figure ~\ref{fig:global optimization}(left)). 
To eliminate these problems, we use a similar method as~\cite{Zhao2018} to optimize the final toolpath, \revision{in which they iteratively evolve a single toolpath considering the spacing and smoothing constraints.}
Different from the original method, our input is not a single path but multiple continuous paths. \autoref{fig:global optimization} shows a result before and after optimization.
% find a satisfying solution

%From the final path visualization, we can observe two adjacent paths may be so close to each other that overlap or so far that generate a gap. This usually happens at the junction of the two OPPs caused by our path planning for every OPP separately, the flatten area and the inter-layer connection path which contains extra path. \llu{this para needs revision. state the reason first, not the vis}Figure \ref{fig:global optimization}(left) shows this phenomenon. 
\section{Results and Discussions}

This section shows OPP decomposition results on \revision{surface} models with varying geometric complexity, printed results, and comparisons with the Ultimaker Cura \revision{4.9.1} software. \revision{We evaluate our ACAP algorithm on DIW-based ceramic printing and FDM platforms, with clay and thermoplastics as the material, respectively.}

\subsection{Implementation and Parameters}

\begin{figure}[tbh]
\centering
\includegraphics[width=1.0\linewidth]{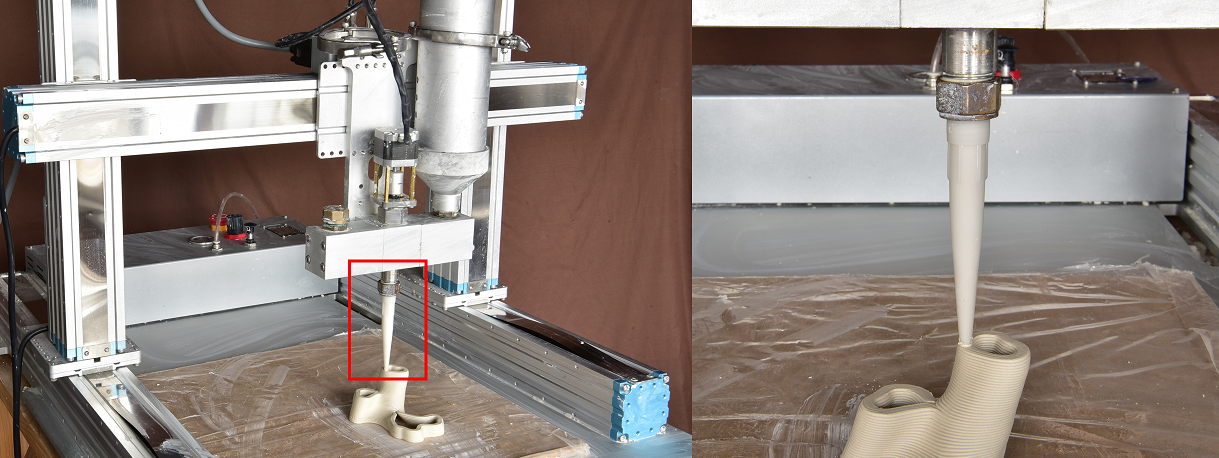}
%\vspace{-8pt}
\caption{The three-axis printer \textit{Eazao Mega 5} for experiments (left) with a slender nozzle (right).}
\label{fig:machine and nozzle}
\end{figure}

Our algorithm is implemented with C++, running on a PC with an Intel Core i7-9700 CPU @ 3.0GHz and 32GB memory. For the printing experiments, we use a 3-axis \revision{DIW} ceramic printer \textit{Eazao Mega 5} with 470×370×390$mm^3$ printing volume (\autoref{fig:machine and nozzle}) and \revision{an FDM printer \textit{Hori Z560} with 360×350×500$mm^3$ printing volume}. 
\revision{Note that the parameters below in the brackets refer to the FDM.}
%This machine relies on an air pump to extrude clay.
\revision{For the ceramic printing (FDM) printer,} we use a 90 (\revision{8})$mm$ long nozzle with \revision{the nozzle diameter} of \revision{5.2 (1.0)}$mm$ (\autoref{fig:machine and nozzle} shows the ceramic printer's nozzle), nozzle movement speed as 25.0 \revision{(25.0)}$mm/s$. The printing path width is \revision{set to 6 (1.5)$mm$.}
\revision{In \autoref{sec:fabcon}, we take the average of the layer thickness range, i.e., $t = $1.5 (0.35)$mm$ for calculating slope angle constraint}, which is 30$^{\circ}$ \revision{(35$^{\circ}$)} according to the formulation.
We set the flat layer thickness to 1.0 \revision{(0.2)}$mm$ for slicing, producing the best surface quality in our experiments, and also use it as the slicing layer thickness to slice models and generate $G_{depend}$.
% During the algorithm, we set $\lambda = 0.8$ for orientation assessment (\autoref{sec:orientation-assessment})\llu{need revision},
The distance threshold $D$ is \revision{5 (2)$mm$} in connecting layers (\autoref{sec:PathGeneration}).

\begin{figure}[tbh]
\centering
\includegraphics[width=1.0\linewidth]{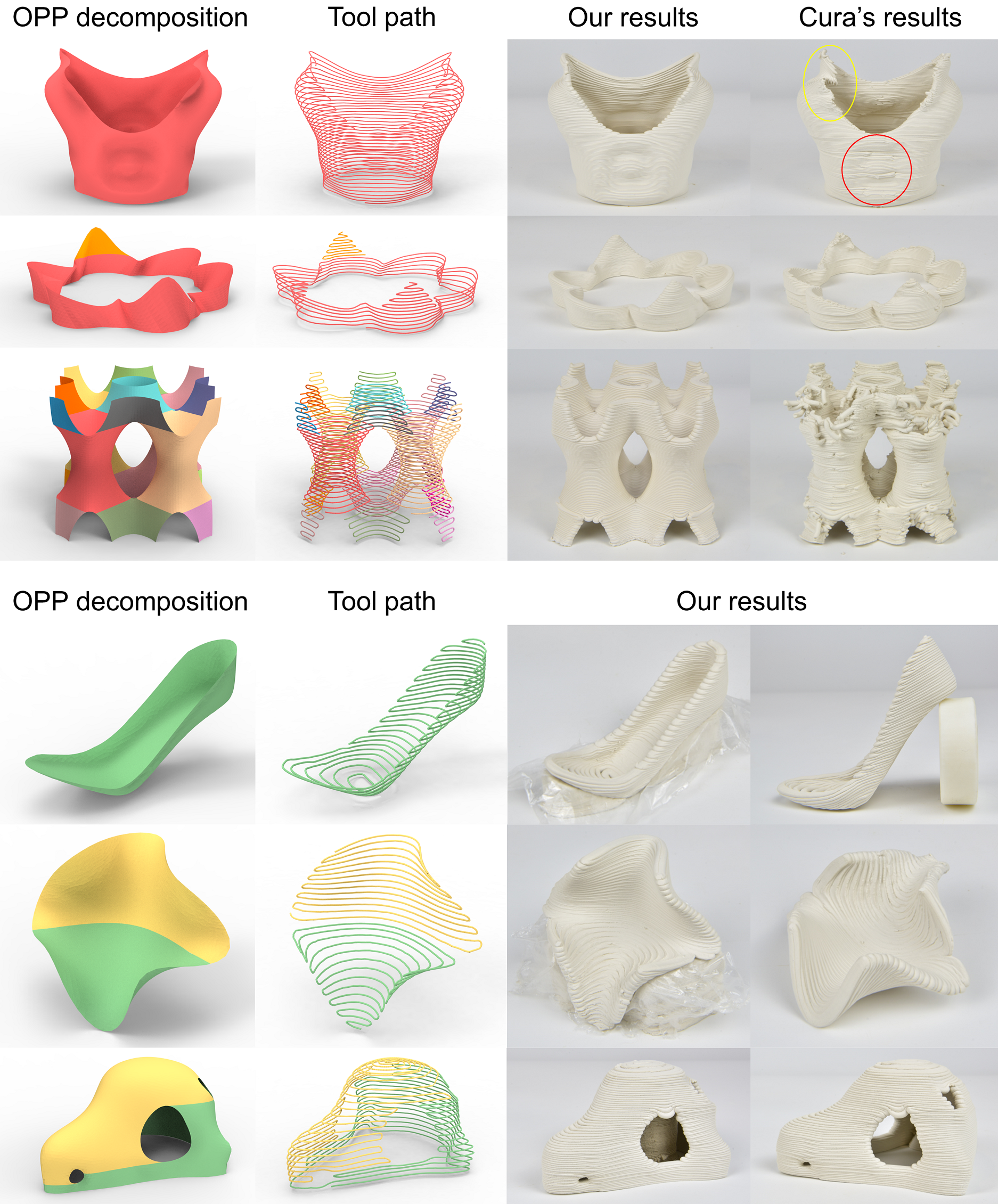}
%\vspace{-7pt}
\caption{\revision{Some ceramic 3D printing results.
The models in the top three rows are self-supporting. The OPP decomposition, toolpath, and printed results by our method and Cura are shown for each row, respectively. 
%Our method outperforms Cura for shell models in surface quality (less artifact and slighter deformation).  
The last three rows show models with support. We pre-build the support structures and insert them manually during fabrication.}}
\label{fig:gallery-1}
\end{figure}

\begin{figure}[tbh]
\centering
\includegraphics[width=1.0\linewidth]{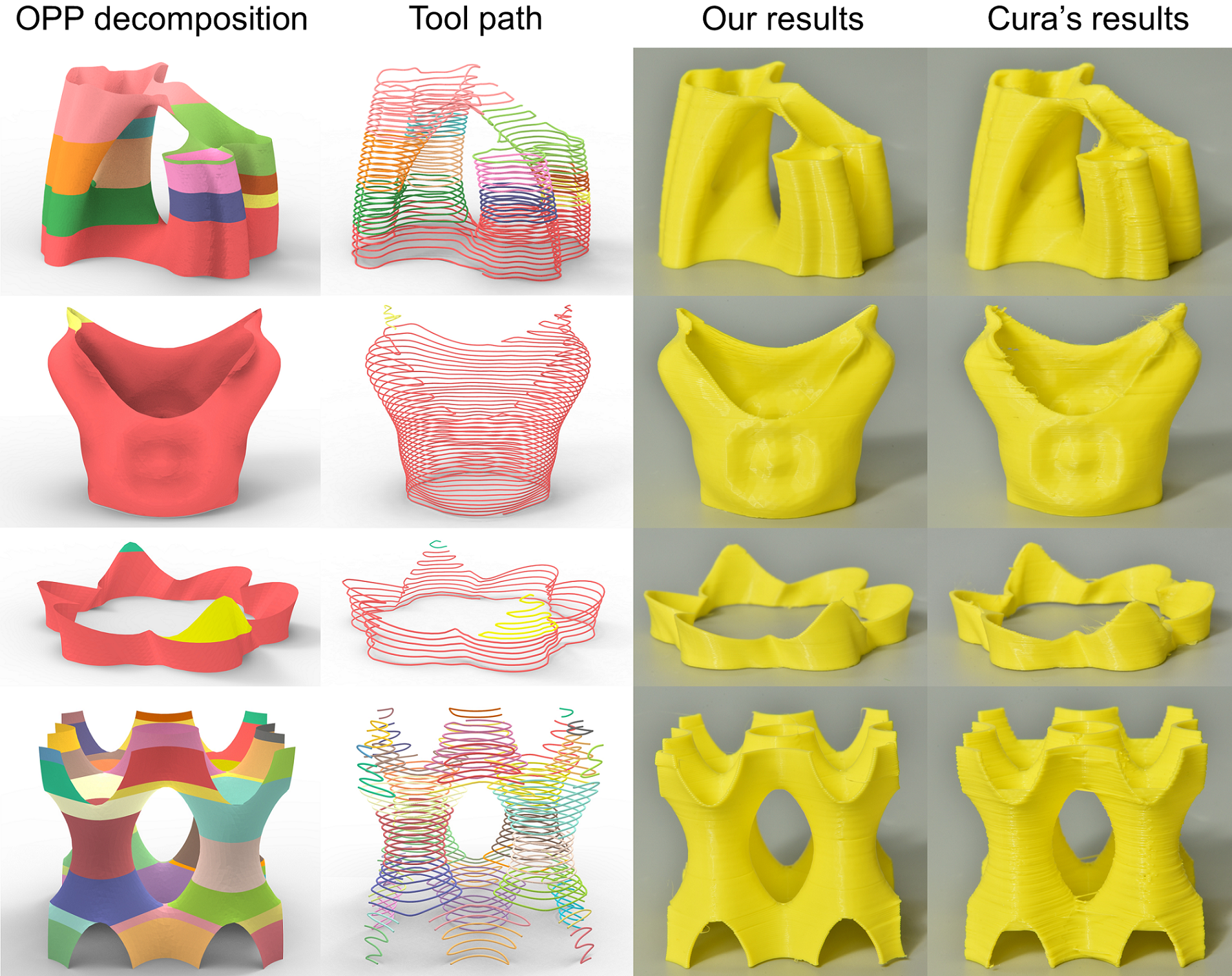}
%\vspace{-8pt}
\caption{\revision{Some FDM printing results. Left-to-right: the OPP decomposition, toolpath, and printed results by our method and Cura, respectively.
Compared with ceramic printing, the FDM results have more OPPs due to the shorter nozzle to avoid collisions. 
The surface quality is better than ceramic printing for both methods, while the quality outperforming is still visible.}}
\label{fig:gallery-3}
\end{figure}

\begin{figure}[tbh]
\centering
\includegraphics[width=1.0\linewidth]{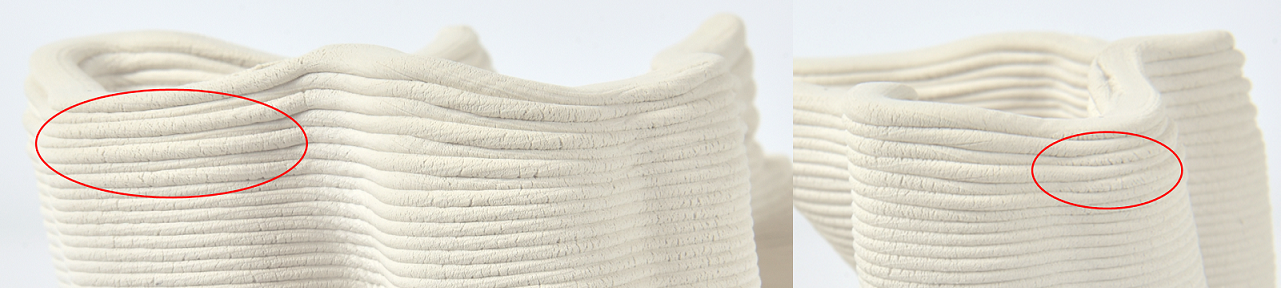}
%\vspace{-8pt}
\caption{\revision{The layer thickness is non-uniform in the outer boundary of curved layers. The main reason is that the material deforms in different ways when extruded in downhill or uphill directions.}}
\label{fig:groove}
\end{figure}

\subsection{Fabrication Results}

\begin{table}[tbh]
\renewcommand{\arraystretch}{1.2}
\centering
\begin{tabular}{m{1.8cm}<{\hfill}| m{0.35cm}<{\centering} m{0.4cm}<{\centering}  m{0.5cm}<{\centering} m{0.5cm}<{\centering} m{0.6cm}<{\centering} m{0.6cm}<{\centering} m{0.6cm}<{\centering}}
\hline
\centering{Model} 
& \centering{H} 
& \centering{\hspace*{-1pt}\#OF}
& \centering{\#OO} 
& \centering{\#OC}  
& {$\rm T_{ours}$} 
& {$\rm T_{cura}$} 
& {$\rm T_{save}$}\\
\hline
%\rule{0pt}{15pt}
{\begin{minipage}[b]{0.06\columnwidth}
		\centering
		\raisebox{-.3\height}{\includegraphics[width=0.8\linewidth]{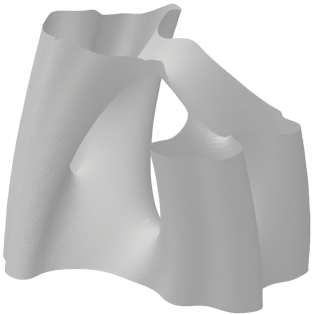}}
	\end{minipage}
Julia vase}     &{81} &{8} &{4} &{\hspace*{2pt}\revision{832}} &{17.7} &{34.3} &{\revision{48\%}} \\
\hline
{\begin{minipage}[m]{0.06\columnwidth}
		\centering
		\raisebox{-.3\height}{\includegraphics[width=0.8\linewidth]{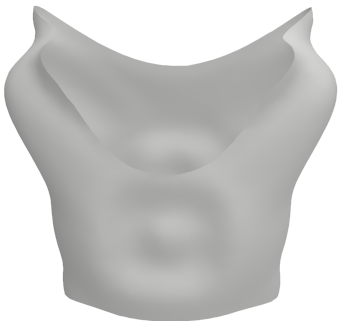}}
	\end{minipage}
Grail}    &{\hspace*{-1pt}120} &{2}  &{1} &{\revision{1541}} &{20.1} &{26.7}  &{\revision{25\%}}         \\
\hline
{\begin{minipage}[b]{0.06\columnwidth}
		\centering
		\raisebox{-.2\height}{\includegraphics[width=0.9\linewidth]{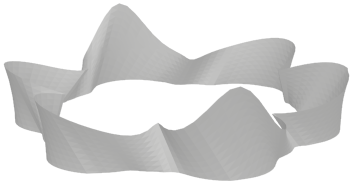}}
	\end{minipage}
Crown}    &{38} &{6}  &{2} &{\revision{\hspace*{2pt}116}} &{8.3} &{12.5}   &{\revision{34\%}}         \\
\hline
{\begin{minipage}[b]{0.06\columnwidth}
		\centering
		\raisebox{-.3\height}{\includegraphics[width=0.8\linewidth]{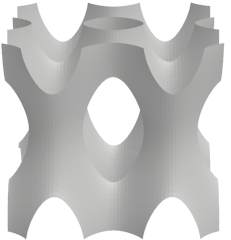}}
	\end{minipage}
TPMS}    &{95} &{18}  &{18} &{\revision{3332}} &{37.3} &{58.5}  &{\revision{36\%}}        \\
\hline
\hline
{\begin{minipage}[b]{0.06\columnwidth}
		\centering
		\raisebox{-.3\height}{\includegraphics[width=0.8\linewidth]{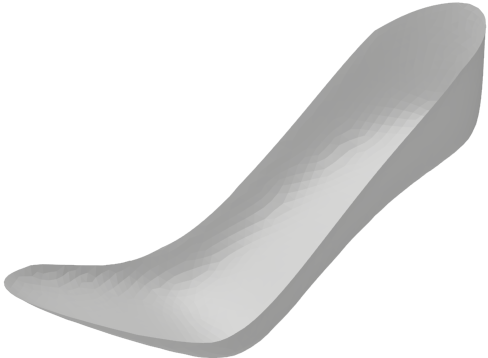}}
	\end{minipage}
Shoe}    &{52} &{2}  &{1} &{{\hspace*{1pt}N/A}} &{20.7} &{{N/A}}   &{\revision{N/A}}          \\
\hline 
{\begin{minipage}[b]{0.06\columnwidth}
		\centering
		\raisebox{-.3\height}{\includegraphics[width=0.8\linewidth]{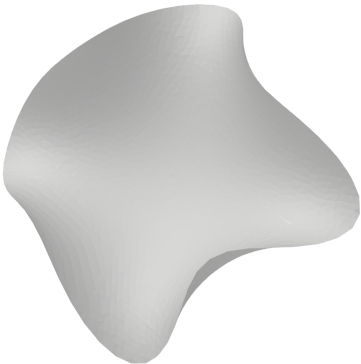}}
	\end{minipage}
Ocean}   &{79} &{4}  &{2} &{{\hspace*{1pt}N/A}} &{59.5} &{{N/A}}&{\revision{N/A}} \\ 
\hline
{\begin{minipage}[b]{0.06\columnwidth}
		\centering
		\raisebox{-.3\height}{\includegraphics[width=0.8\linewidth]{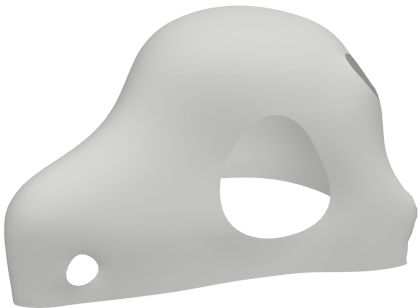}}
	\end{minipage}
Mask}  &{54} &{6}  &{2} &{\hspace*{1pt}N/A} &{8.6} &{N/A}    &{\revision{N/A}}     \\
\hline
\hline
\revision{{\begin{minipage}[b]{0.06\columnwidth}
		\centering
		\raisebox{-.3\height}{\includegraphics[width=0.8\linewidth]{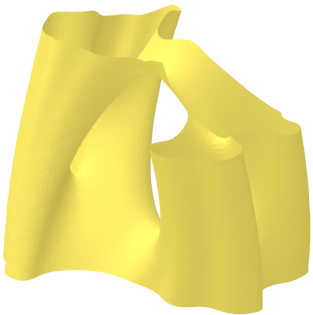}}
	\end{minipage}
Julia vase}}     &{\revision{41}} &{\revision{15}}  &{\revision{11}} &{\revision{\hspace*{2pt}741}} &{\revision{19.8}} &{\revision{35.0}}  &{\revision{43\%}}\\
\hline
\revision{{\begin{minipage}[m]{0.06\columnwidth}
		\centering
		\raisebox{-.3\height}{\includegraphics[width=0.8\linewidth]{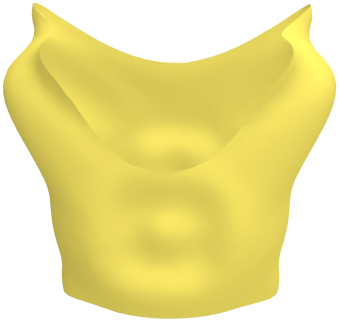}}
	\end{minipage}
Grail}}    &{\revision{60}} &{\revision{3}}  &{\revision{2}} &{\revision{1487}} &{\revision{24.0}} &{\revision{31.4}} &{\revision{24\%}}    \\
\hline
\revision{{\begin{minipage}[b]{0.06\columnwidth}
		\centering
		\raisebox{-.2\height}{\includegraphics[width=0.9\linewidth]{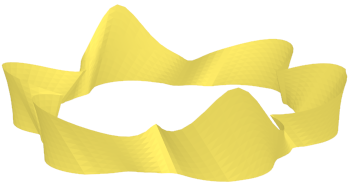}}
	\end{minipage}
Crown}}    &{\revision{19}} &{\revision{7}} &{\revision{3}} &{\hspace*{2pt}\revision{311}} &{\revision{8.8}} &{\revision{12.7}} &{\revision{31\%}}     \\
\hline
\revision{{\begin{minipage}[b]{0.06\columnwidth}
		\centering
		\raisebox{-.3\height}{\includegraphics[width=0.8\linewidth]{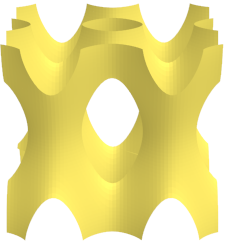}}
	\end{minipage}
TPMS}}    &{\revision{48}} &{\revision{44}} &{\revision{44}} &{\revision{2971}} &{\revision{34.5}} &{\revision{51.4}}  &{\revision{33\%}}        \\
\hline

\end{tabular}
\caption{Statistics of the results. 
\revision{The gray icons represent the ceramics printing results (\autoref{fig:teaser} and \autoref{fig:gallery-1}), and the yellow icons represent the FDM printing results (\autoref{fig:gallery-3})}.
H is the model height (mm). 
\#OF is the number of OPPs of our method with only flat layers. 
\#OO is the number of OPPs using both flat and curved layers. 
\#OC is the number of OPPs for Cura. 
\revision{
$\rm T_{ours}$ and $\rm T_{cura}$ indicate the printing time (minutes) of our method and Cura, respectively.
$\rm T_{save}$ is the percentage of time saved by our method compared to Cura. 
For the three models \textit{Shoe, Ocean, Mask}, we did not perform Cura as they need support structures.}}
\label{table:data}
\end{table}

% %%%%%describe teaser result first%%%%%
% \autoref{fig:teaser} shows the printed \textit{Julia vase} model \revision{printed by the  ceramic printer}, with our method and Cura. 
% There are 4 separated columns in the middle part of the model \hs{what is middle part?}, and thus Cura failed to spiralize the multiple-contour layers, producing 195 disconnected printing tool paths, and our method decomposes the model into 4 OPPs, generates the ACAP tool path along both flat and curved slicing layers, thus achieving superior printing quality and reducing the printing time by almost half. \hs{we have a clear caption to explain this result. Right? Why do we need to do it again?}

\autoref{fig:teaser}, \autoref{fig:gallery-1} and \autoref{fig:gallery-3} show results \revision{of two different printing processes (ceramic printing and FDM)}, including OPP decomposition, printing path, and printed models (using our method and Cura, respectively). 
Refer to \autoref{table:data} for the statistics, \revision{where} we list each model's height, the number of OPPs \revision{produced by our algorithm, and the printing time raised by our method and Cura software. Specifically, the table includes} the number of OPPs after merging through flat layers (\#OF), \revision{and the number of OPPs after merging through curved layers (\#OO),} to illustrate the effectiveness of \textit{stacking} and \textit{curving} operations. 
\revision{Note that we scale models to half size in FDM experiments for time saving.
Since the nozzle used for FDM printing is much shorter than for ceramic printing, our method partitions more OPPs to avoid the global collisions.
}

%%%%%compare with Cura, and clarify the reasons%%%%%
\paragraph{Compare with Cura}
\revision{The top three models in \autoref{fig:gallery-1} and the four models in \autoref{fig:gallery-3} are all self-supporting. In the setting of Cura, we choose the \textit{Surface Mode}, enable \textit{Spiralize Outer Contour} and \textit{Retraction}, and use the same nozzle movement speed as our method.}
It shows that our method outperforms Cura for shell models containing multiple \textit{contours} or \textit{segments} in both \revision{surface quality (fewer artifacts and slighter deformation) and fabrication efficiency (save 24\%$\thicksim$48\% of printing time)}.
\revision{For example, as the \textit{Grail} model printed with Cura (ceramic printing), a large number of transfer moves (\revision{1541}) induce deformation (see the red circle in \autoref{fig:gallery-1}).}
\revision{Such transfer moves would increase the forces on the printed part and may cause the collapse (see the yellow circle).
In contrast, the ACAP toolpath for this model only has one transfer move in FDM, and even no transfer move in ceramic printing, and thus strengthens the model during fabrication and avoids these problems. 
Generally, our algorithm improves the printing efficiency more significantly for the models with multiple branches like the \textit{Julia vase} or \textit{TPMS} model. 
%
%shows a more obvious comparison, as severe artifacts can be observed for the results from Cura.
Moreover, we can see the benefit of curved layers, which reduces the number of OPPs by around 50\%, and improves the staircase defect (see the upper part of the \textit{Crown} model).
The \textit{TPMS} model has no curved layer, as the slope angle constraint restricts it.
}

%\paragraph{Models with support structure}
\revision{
For the models that require support structures, we pre-build the support structures and place them before printing the models. See the last three models in \autoref{fig:gallery-1}}.
To make support structure easy to remove, we add a membrane on the surface of their contact before printing the model.
%Note that Cura cannot generate support for shell surfaces and is thus invalid for these models.

%%%%%clarify some drawbacks: seam-line, over-extrusion, sagging,hardware-related factors%%%%%
\revision{
\paragraph{Discussions on the printing quality}
%common defects
Generally, the quality of ceramic printing is more sensitive to toolpath continuity for surface models than FDM printing. 
Therefore, the clay results from the ACAP algorithm show significant superiority in the model quality over those from Cura.
However, we can still observe some visible artifacts in the clay printouts, like over-extrusion at the zig-zag turning areas, seams at the junction of OPPs, and sagging at the concavity corner.
The primary reason is that the DIW-based ceramic printing technique with clay is less mature than FDM with thermoplastics.
The extrusion amount can hardly be precisely controlled due to material inertia.
Moreover, there are many uncertainties during the fabrication process, such as material humidity, air pressure, and other coupled hardware and material problems.
}

\revision{
We also remark a defect in curved layers. See \autoref{fig:groove}; we observe that the layer thickness is non-uniform in the outer boundary, appearing like the stair-case defect.
The reason is that curved slicing layers are applied with adaptive layer thickness, and the material deforms in different behaviors when extruded in downhill or uphill directions.
Such artifact might be improved via fine-tuning the material flow or the nozzle height, while highly demanding on three issues: stable properties of the clay material, accurate simulation of clay deformation, and precise control of the extrusion rate for clay in the printing platform. 
}

\subsection{Algorithm Performance}
\begin{table}[tb]
\small 
\revision{
\renewcommand{\arraystretch}{1}
\begin{tabular}{m{0.4cm}<{\centering}|
m{0.2cm}<{\centering} m{0.3cm}<{\centering} >{\columncolor{mygray}}m{0.26cm}<{\centering}| 
m{0.2cm}<{\centering} m{0.2cm}<{\centering} m{0.2cm}<{\centering} >{\columncolor{mygray}}m{0.26cm}<{\centering}|
m{0.2cm}<{\centering} m{0.25cm}<{\centering}  >{\columncolor{mygray}}m{0.26cm}<{\centering}|
>{\columncolor{mygray}}m{0.25cm}<{\centering}|
>{\columncolor{mygray}}m{0.25cm}<{\centering}|
>{\columncolor{mygray}}m{0.34cm}<{\centering}}  
\hline
\rowcolor{white}{}&\multicolumn{3}{c|}{Section 4.1}&\multicolumn{4}{c|}{Section 4.2}&\multicolumn{3}{c|}{Section 4.3} &{\hspace*{-1pt}4.4}&\multicolumn{1}{c|}{5}&\multicolumn{1}{c}{All}\\
% \multicolumn{16}{|c|}{3} 
\hline
\centering{{\hspace*{-5pt}Model}}&{\hspace*{-3pt}\#DN}&{\hspace*{-3pt}\#DE} &{$\rm T_d$} &{\hspace*{-3pt}\#IN} &{\hspace*{-3pt}\#IE} &{\hspace*{-4pt}\#BS}&{$\rm T_f$} &{\hspace*{-3pt}\#SO} &{\hspace*{-4pt}\#CU} &{$\rm T_c$} &{$\rm T_l$}  &{$\rm T_o$} &{$\rm T_t$}\\
\hline
{\begin{minipage}[b]{0.045\columnwidth}
        \centering
		\raisebox{-.3\height}{\includegraphics[width=0.8\linewidth]{images/juliaVaseIco.png}}
 \end{minipage}}&{81} &{\hspace*{-2pt}209}&{$\epsilon$} &{\hspace*{-1pt}17} &{\hspace*{-2pt}19} &{\hspace*{-1pt}8} &{$\epsilon$} &{\hspace*{-2pt}128}&{\hspace*{-2pt}768} &{\hspace*{-1pt}315}&{\hspace*{-2pt}246} &{\hspace*{-2pt}326} &{\hspace*{-2pt}887}   \\ 
\hline
{\begin{minipage}[b]{0.045\columnwidth}
		\centering
		\raisebox{-.3\height}{\includegraphics[width=0.8\linewidth]{images/grailIco.png}}
	\end{minipage}
}&{\hspace*{-2pt}120}&{\hspace*{-2pt}146}    &{$\epsilon$} &{\hspace*{-1pt}4} &{\hspace*{-2pt}3} &{\hspace*{-1pt}2}&{$\epsilon$} &{2}&{4}  &{3} &{\hspace*{-2pt}497}  &{\hspace*{-2pt}438} &{\hspace*{-2pt}938} \\  
\hline
{\begin{minipage}[b]{0.05\columnwidth}
		\centering
		\raisebox{-.2\height}{\includegraphics[width=0.9\linewidth]{images/crownIco.png}}
	\end{minipage}
} &{38}&{\hspace*{-2pt}91}   &{$\epsilon$} &{\hspace*{-1pt}9}&{\hspace*{-2pt}8} &{\hspace*{-1pt}6} &{$\epsilon$} &{\hspace*{-1pt}16} &{96}  &{39} &{\hspace*{-2pt}313} &{\hspace*{-2pt}144} &{\hspace*{-2pt}496}    \\ 
\hline
{\begin{minipage}[b]{0.045\columnwidth}
		\centering
		\raisebox{-.3\height}{\includegraphics[width=0.8\linewidth]{images/tpmsIco.png}}
	\end{minipage}
}  &{95}&{\hspace*{-2pt}598}  &{$\epsilon$}&{\hspace*{-1pt}38} &{\hspace*{-2pt}48} &{\hspace*{-1pt}18} &{3} &{\hspace*{-1pt}96}&{0} &{3}&{1}  &{\hspace*{-2pt}455} &{\hspace*{-2pt}462} \\   
\hline
\hline
{\begin{minipage}[b]{0.05\columnwidth}
		\centering
		\raisebox{-.3\height}{\includegraphics[width=0.8\linewidth]{images/shoeIco.png}}
	\end{minipage}
}  &{52}&{\hspace*{-2pt}64}  &{$\epsilon$} &{\hspace*{-1pt}3} &{\hspace*{-2pt}2}  &{\hspace*{-1pt}2} &{$\epsilon$} &{2}&{2}  &{2}&{84} &{53} &{\hspace*{-2pt}139}       \\
\hline
{\begin{minipage}[b]{0.045\columnwidth}
		\centering
		\raisebox{-.3\height}{\includegraphics[width=0.8\linewidth]{images/oceanIco.png}}
	\end{minipage}
}&{79}&{\hspace*{-2pt}97} &{$\epsilon$} &{\hspace*{-1pt}7} &{\hspace*{-2pt}9} &{\hspace*{-1pt}4} &{$\epsilon$} &{2} &{8} &{7}&{\hspace*{-2pt}419} &{94} &{\hspace*{-2pt}520} \\
\hline
{\begin{minipage}[b]{0.05\columnwidth}
		\centering
		\raisebox{-.3\height}{\includegraphics[width=0.8\linewidth]{images/caraIco.png}}
	\end{minipage}
} &{57}&{\hspace*{-2pt}122} &{$\epsilon$} &{\hspace*{-1pt}12}&{\hspace*{-2pt}15} &{\hspace*{-1pt}6} &{$\epsilon$} &{\hspace*{-2pt}121} &{\hspace*{-2pt}741} &{\hspace*{-1pt}274}&{\hspace*{-2pt}290}&{\hspace*{-2pt}242} &{\hspace*{-2pt}806}  \\       
\hline
\hline
{\begin{minipage}[b]{0.045\columnwidth}
        \centering
		\raisebox{-.3\height}{\includegraphics[width=0.8\linewidth]{images/juliaVaseIco2.png}}
 \end{minipage}}&{\hspace*{-2pt}205}&{\hspace*{-3pt}1254}&{$\epsilon$}&{\hspace*{-2pt}443} &{\hspace*{-4pt}1182} &{\hspace*{-1pt}15} &{3} &{\hspace*{-1pt}48} &{\hspace*{-2pt}288}  &{\hspace*{-1pt}127}&{\hspace*{-2pt}274} &{\hspace*{-2pt}495} &{\hspace*{-2pt}899}   \\ 
\hline
{\begin{minipage}[b]{0.045\columnwidth}
		\centering
		\raisebox{-.3\height}{\includegraphics[width=0.8\linewidth]{images/grailIco2.png}}
	\end{minipage}
} &{\hspace*{-2pt}300}&{\hspace*{-2pt}415}   &{$\epsilon$}&{\hspace*{-2pt}107} &{\hspace*{-3pt}157} &{\hspace*{-1pt}3} &{$\epsilon$} &{2}&{2}  &{1}&{\hspace*{-2pt}440} &{\hspace*{-2pt}524} &{\hspace*{-2pt}965} \\  
\hline
{\begin{minipage}[b]{0.05\columnwidth}
		\centering
		\raisebox{-.2\height}{\includegraphics[width=0.9\linewidth]{images/crownIco2.png}}
	\end{minipage}
} &{\hspace*{-2pt}190}&{\hspace*{-3pt}1043}   &{$\epsilon$}&{\hspace*{-2pt}384} &{\hspace*{-3pt}971} &{\hspace*{-1pt}7} &{$\epsilon$} &{\hspace*{-1pt}28} &{\hspace*{-2pt}196} &{\hspace*{-1pt}131} &\hspace*{-2pt}{298}  &{\hspace*{-2pt}257}  &{\hspace*{-2pt}686}   \\ 
\hline
{\begin{minipage}[b]{0.045\columnwidth}
		\centering
		\raisebox{-.3\height}{\includegraphics[width=0.8\linewidth]{images/tpmsIco2.png}}
	\end{minipage}
} &{\hspace*{-2pt}240}&{\hspace*{-3pt}3836}   &{$1$} &{\hspace*{-2pt}905}&{\hspace*{-4pt}3295} &{\hspace*{-1pt}44} &{37}&{1} &{0} &{$\epsilon$} &{1} &{\hspace*{-2pt}573} &{\hspace*{-2pt}612} \\   
\hline
\end{tabular}
\caption{
\revision{Statistics of the algorithm performance. "\#" indicates amount, "T" indicates running time (seconds). "$\epsilon$" indicates the running time is negligible (<0.1s).
Specifically, the table shows the number of nodes (\#DN), edges in $G_{depend}$ (\#DE, including dependency edges and collision dependency edges), nodes in $G_{init}$ (\#IN), and edges in $G_{init}$ (\#IE), the depth of beam search (\#BS), the number of optimal solutions of beam search (\#SO) and calling CurviSlicer without smooth term (\#CU).
For the running time, we shows the time of building $G_{depend}$ ($\rm T_d$) and $G_{flat}$ ($\rm T_f$), running time for merging $G_{curved}$ ($\rm T_c$), performing CurviSlicer with smooth term and the layers connection ($\rm T_l$), finishing toolpath optimization ($\rm T_o$), and the total running time ($\rm T_t$).}
}
\label{table:timing}
}
\end{table}

\begin{figure}[tbh]   
\centering
\includegraphics[width=1.0\linewidth]{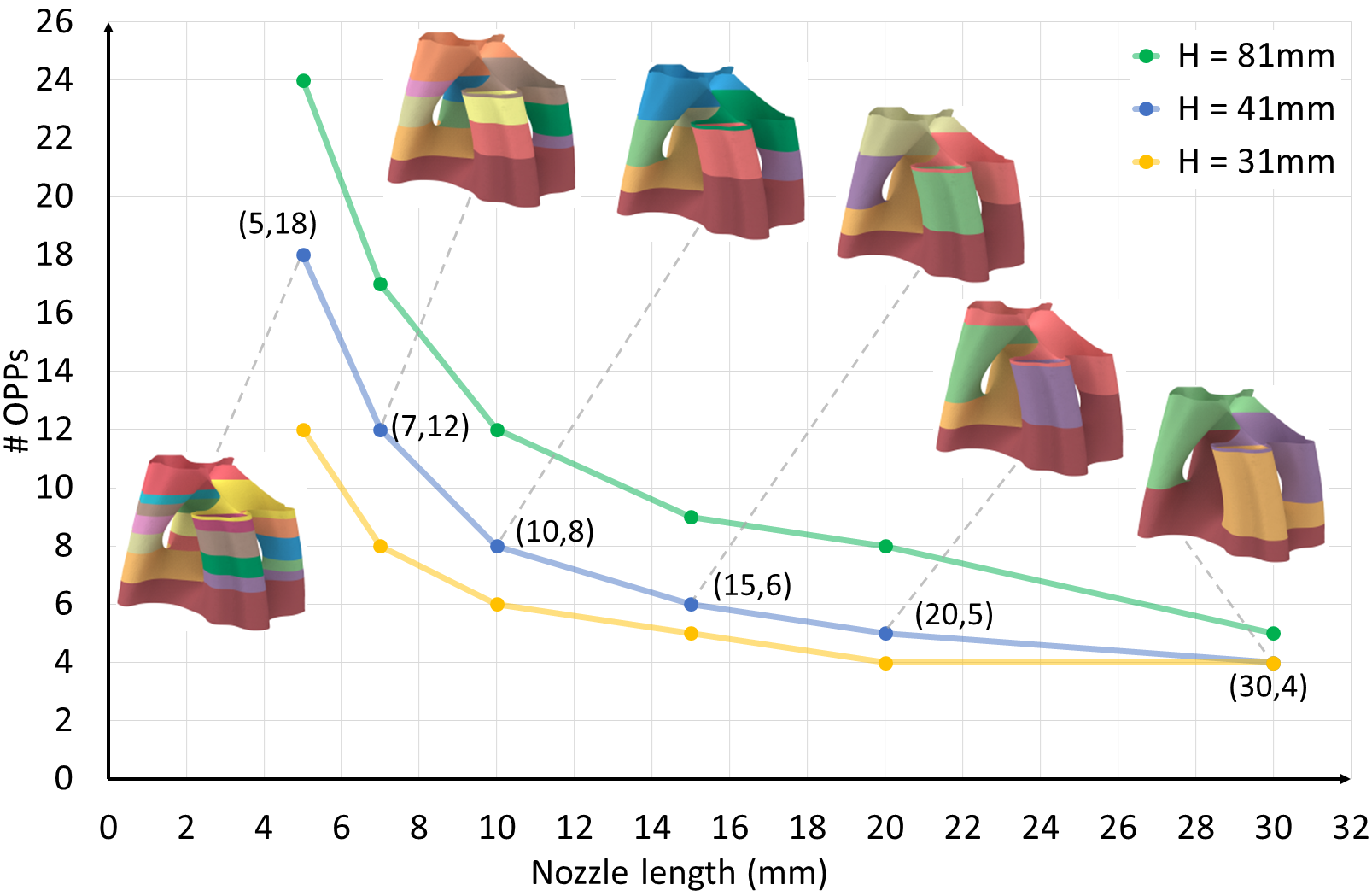}
\caption{\revision{The decomposed OPP numbers regarding different model sizes (H indicates the height) and nozzle lengths on the \textit{Julia vase} model in the FDM printing setup.
With the nozzle length increasing, the number of OPPs decreases significantly as the global collision happens less.}}
\label{fig:DifferentNozzles}
\end{figure}

\revision{
\paragraph{Effect of nozzle length}
The nozzle size, especially the nozzle length, also affects the number of OPPs, as the shorter nozzle indicates more possibilities for the collision between the model and the printing platform. 
See \autoref{fig:DifferentNozzles}, we test the nozzle length from 5mm to 30mm on the \textit{Julia vase} model in three different sizes.
}
%when we use a long nozzle (30mm), there is no collision dependency edge, and  is finally decomposed to 4 OPPs. In short, our algorithm performs better with long nozzle. 

% Improve 
% The major computational cost of our framework includes \revision{collision detection}, OPP merging, and toolpath optimization. 
\revision{
\paragraph{Running time}
\autoref{table:timing} shows the running time statistics with the number of nodes and edges of the OPP graph during each step of our ACAP algorithm.
For most of the input models, our algorithm is extremely efficient in the generation of $G_{depend}$, $G_{init}$ and $G_{flat}$, since our algorithm does not involve any geometric computation in this phase, but only some graph operations. 
The only exception is the \textit{TPMS} model (FDM),
whose time of building $G_{flat}$ is much longer than that of other models, since it owns a large number of nodes and edges in $G_{init}$ (905 and 3295) with more iterations of beam search (44). The running time of OPP merging through curved layers varies mainly according to the number of $G_{flat}$ (\#SO) and calls to CurviSlicer (\#CU), where CurviSlicer is the bottleneck to speed up this step. Similarly, the time of layers connection ($\rm T_l$) is also time-consuming, which would apply CurviSlicer with the smooth term for each decomposed OPP patch. Besides, the toolpath optimization also takes a long time, worth improving.}

\section{Conclusion, Limitation, and Future work}
\label{sec:conclusion}

In extrusion-based 3D printing, path continuity significantly impacts the surface quality and printing time, especially for shell models. We put forward the original concept of OPP to quantify path continuity and propose a method to decompose the given shell model into as few OPPs as possible, considering manufacturing constraints on a standard three-axis printer platform. We demonstrated our methods on various models, and the results are superior to existing methods in surface finish and printing time.

\begin{figure}[tb]
\centering
\includegraphics[width=1.0\linewidth]{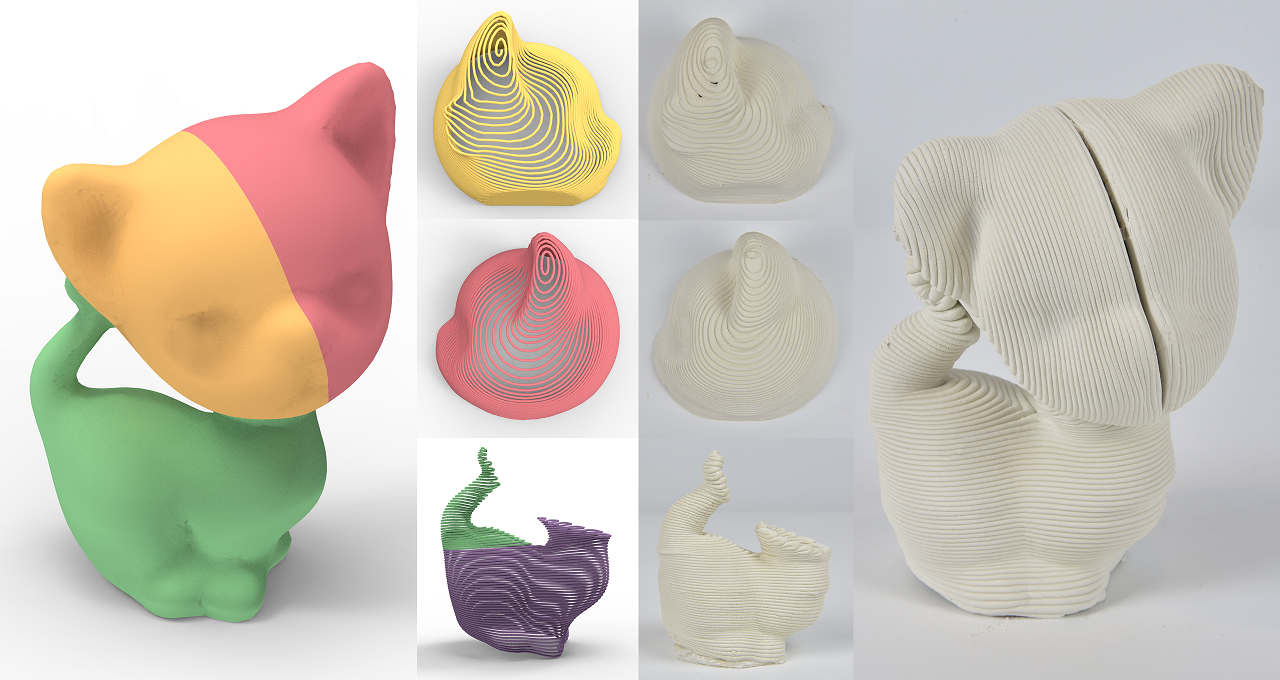}
\caption{For the kitten model that does not satisfy the support structure constraint, we manually decompose it into three patches satisfying the constraint (left), separately print them using our ACAP toolpaths (middle), and assemble them afterwards (right).}
\label{fig:kitten}
\end{figure}

\begin{figure}[tb]
\centering
\includegraphics[width=1.0\linewidth]{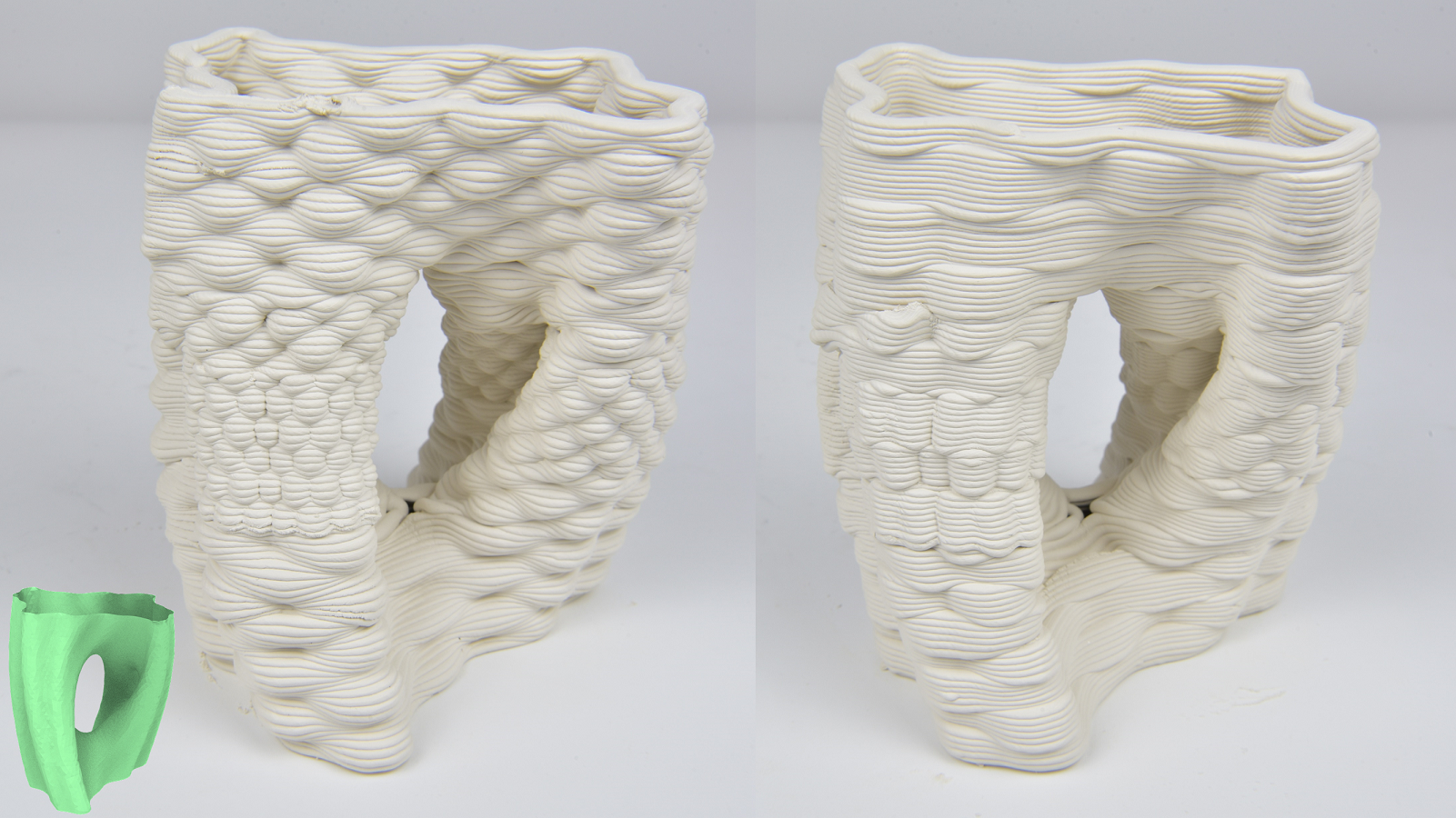}
\caption{Julia vase model is decomposed into four OPPs, embedded with different textures.}
\label{fig:texture}
\end{figure}

\paragraph{Limitations and Future work}
\revision{
Regarding the toolpath, improvement space still exists.
The inter-layer connecting path might be outside the model by a distance $t_{min}$, and such an artifact cannot be guaranteed to be avoided through the post global toolpath optimization.
The constant path width maintains the uniform horizontal wall thickness of the shell model; however, the shell thickness in the direction normal to the surface varies, which is an unintended consequence.
}

\revision{
Our framework only allows for geometry to be printed which is surface model. Extending the OPP criterion and the ACAP algorithm to general solid models could be valuable continuous work.
The key issue is how to adapt the OPP criterion to different interior structures or infilling patterns.
Moreover, to fully explore the applications of surface models, applying the shell reinforcement techniques like adding ribs~\cite{GilUreta2020} or modulating the thickness~\cite{Xing2021} is also an exciting direction. In such a way, the contour of the surface models is of varying thickness. The ACAP algorithm should be adapted to combine both thin regions of +/- the nozzle size and wider ranges of multiple nozzle sizes.
}

% \revision{Considering the instability during ceramic printing process since the extruding clay cannot be solidification immediately to support the upper layers}, \cite{Xing2021} proposed to locally enhance the structural stability by shell thickening for self-supported models.
% \revision{But} it's not straightforward to extend their method to shell models with support structures.
%
%We would like to study the adaptions of OPP on surface models with multi-layers, solid models with different infills as the continuous work.
%It cannot be directly used in general solid model. Extending our OPP definition and decomposition methods to solid model could be a valuable future work.

\revision{
The support structure constraint limits the feasible surface shapes for fabrication, i.e., the model is either self-supporting or with the support structures located on the ground.
Thus, for models with no orientation satisfying the support structure constraint, we decompose the model manually and assemble them afterwards, as shown in~\autoref{fig:kitten}.
We remark that different degrees of shrinkage of material may leave noticeable seam lines between parts after solidifying of materiel. 
In future work, we would like to study the generalization of more complex surface models by considering more decoupling strategies between the support and intact structures, and also the scheduling strategies for printing the support structures in-situ with the intact model.
An automated pre-decomposition that balances the number of decomposed patches and path continuity is also a natural objective.
%We regard this as an interesting but orthogonal problem for this paper. 
}

\revision{
Since our method generates the toolpath and related G-code files, it would be natural to cooperate with the research works that fine-tune the printing parameters like the extruder height and extrusion amount~\cite{Takahashi2017}, or positions in a local range~\cite{Yan2021}, to achieve various delicate geometric features without violating the toolpath continuity. Referring to~\autoref{fig:texture}, each OPP can be treated as an independent unit to embed different textures on the fly.}

\revision{
Finally, another problem worth exploring is extending the algorithm to multi-axis printing setups. Accordingly, the slope angle constraint can be ignored, the support structures are of fewer restrictions, and the toolpath has more freedom. We believe the research direction owns great potential.
}

\bibliographystyle{ACM-Reference-Format}
\bibliography{ceramics} 
\appendix
\appendix

\section{PC-MPC to PC-TSP}
\label{conversion}
%Given a directed complete graph $G(V,E)$, a distance $D_{ij}$ on each arc $(i,j) \in E $, precedence constraints $\prec$ on $V$, find a minimum distance tour that starts node $1 \in V$, visits all the nodes in $V - \{1\}$, and returns node 1 again so that node i is visited before node j when  $i \prec j$.
The precedence constrained traveling salesman problem (PC-TSP) was formulated by~\cite{Kubo1991} and proved to be NP-hard: "\textit{Given a directed complete graph $G(V,E)$, a distance $D_{ij}$ on each arc $(i,j) \in E $, precedence constraints $\prec$ on $V$, find a minimum distance tour that starts node $1 \in V$, visits all the nodes in $V - \{1\}$, and returns node 1 again so that node i is visited before node j when  $i \prec j$.}" %\cite{Kubo1991} has proved PC-TSP is NP-hard. 
We define our precedence constrained minimum path cover problem (PC-MPC): given a directed acyclic graph $G(V,E)$ where edges are precedence constraints $\prec$ on $V$, our goal is to find a path cover for $G$ with the fewest paths. 

To prove that PC-MPC is NP-hard, we reduce this problem to PC-TSP 
for the input of PC-MPC $G$ to turn it into a weighted directed complete graph: 
(1) set the weight of directed edges of $G$ to $0$. 
(2) add missing edges to make $G$ a directed complete graph $G_{com}$ where the weight of new edges is set to $1$. Therefore, we convert the input of PC-MPC \{$G,\prec$\} to the input of PC-TSP \{$G_{com},D,\prec$\} in linear time. 
The goals of PC-TSP and PC-MPC are equivalent by this conversion. With an arbitrary starting node meeting precedence constraints, find a tour from this node by solving PC-TSP. 
The output of PC-TSP can be taken as the output of PC-MPC by removing edges with a weight of $1$ on tour. The minimum distance of PC-TSP plus equals the number of paths for the minimum path cover of PC-MPC.

\revision{
\section{Orientation Determination}
\label{sec:oridetail}
Given the input model $M$, this step extracts feasible printing orientations by uniformly sampling orientations over the Gaussian sphere, and then choosing one that satisfies the \textit{support structure constraint}.
%See more details about two criteria in \autoref{sec:oridetail}
For each extracted orientation, apply a similar method as~\cite{Hergel2019} to detect the support areas and validate the \textit{support structure constraint}.
First, slice the model and if a layer sampling point requires support, cast a ray downward.
%Suppose the first intersection point lies on the printing platform, add support points for the previous slicing layers below. 
If intersect with the model itself (except adjacent layer), it indicates violating the \textit{support structure constraint}.}

%Finally extract contour-parallel toolpaths inside of contours to generate support structure. 

%which aims to minimize the low slope areas for reducing the under-filling realized by slicing process, and minimize the support structure volume for less material cost and printing time.

%\begin{equation}
%x_{\_normal} = \sum_{i=1}^nw_{i} \cdot bdry_{\_normal}
%\end{equation}

%\begin{equation}
%loss \mathrel{+}= df(bdry) \cdot bdry_{\_normal}  
%\end{equation}

%\begin{equation}
%loss \mathrel{+}= df(x) \cdot x_{\_normal}
%\end{equation}

\revision{
\section{CurviSlicer for Surface Models}
\label{sec:tricks}
CurviSlicer takes the watertight triangle mesh, its tetrahedral mesh, and target flat areas as input, which cannot be directly applied to the surface model, a surface patch of watertight triangle mesh. 
The key challenge is that it's not straightforward to extend the gradient formulation of the vertical coordinates within each tetrahedron (Sec. 4.2 of~\cite{Etienne2019}) to that formulation within each triangle.}

\revision{
We propose to generate a tetrahedral mesh to approximate the original surface patch of watertight triangle mesh with a minimal shell thickness. In our implementation, we set it to $0.3\%$ of the longest diagonal of the bounding box of the input model.
We do not suggest directly offsetting the input surface model along the horizontal direction to form a watertight mesh and generate the corresponding tetrahedral mesh. The self-intersection problem raised by offsetting makes it hard for tetrahedralization. Our solution is described below.
For each triangle, offset its centroid by a minimal distance along the triangle's normal direction, then connect the resulting point with the three points of the triangle to form a tetrahedron.
For each edge of the target flat boundaries (a set of edges of triangles), offset its midpoint by a minimal distance along the horizontal direction starting, then connect the resulting point with the two endpoints to form a triangle.}

\section{Different Merging Results by Curving}
\label{sec:Different merging results}

\begin{figure}[htb]
\centering
\includegraphics[width=1.0\linewidth]{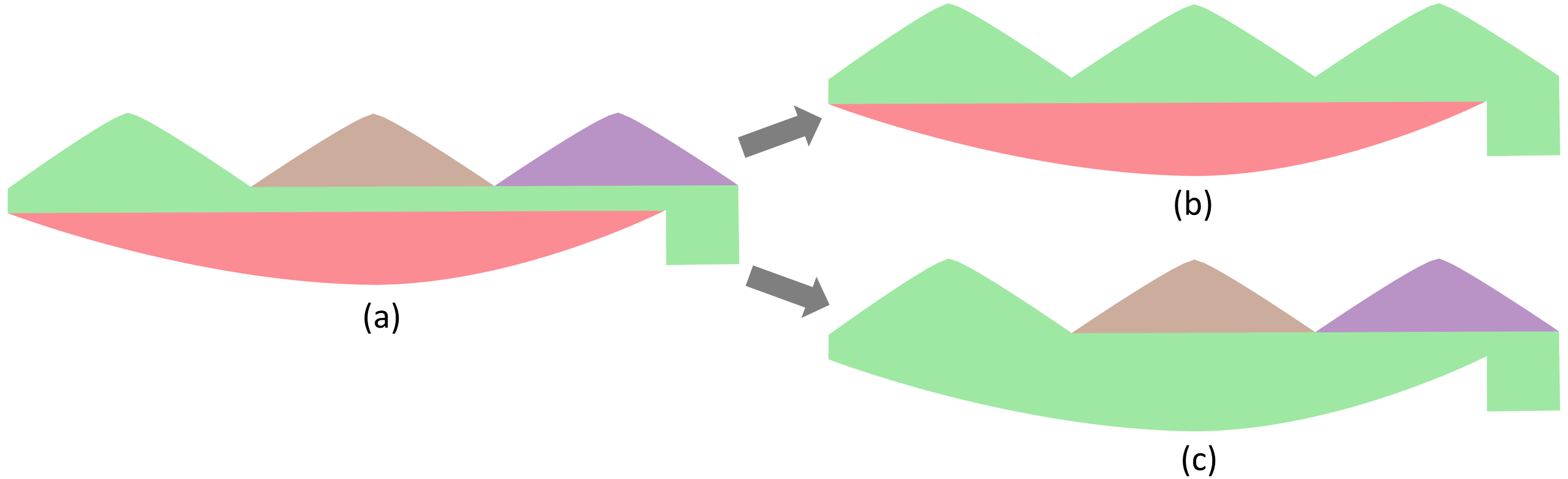}
\caption{\revision{Four I-OPPs can be decomposed by solving PC-MPC (a). If give priority to $curving$ green, brown and purple OPPs, they can merge in sequence and finally merge into two OPPs (b). However, if firstly $curving$ green and red OPPs, the remaining OPPs cannot merge due to the layer thickness constraint, and the final number of OPPs is three (c).}}
\label{fig:failure case 1}
\end{figure}
\revision{
When choosing two OPPs $curving$, different orders may lead to different results. \autoref{fig:failure case 1} is an extreme example. The middle green OPP of (a) meets the layer thickness constraint when $curving$ with only the upper or lower OPP. However, merging with both upper and lower OPPs will violate the constraint. The two different orders of OPP $curving$ (b, c) result in a different number of final OPPs.
}

\section{The Pseudo-code}
\label{sec:pseudocode}
\renewcommand{\algorithmicrequire}{\textbf{Input:}}  % Use Input in the format of Algorithm
\renewcommand{\algorithmicensure}{\textbf{Output:}} % Use Output in the format of Algorithm
    \begin{algorithm}[t]
    \caption{\revision{OPP Merging through Flat Layers}}
    \label{alg:three}
    \begin{algorithmic}[1]
    %\REQUIRE {A directed acyclic graph G(V,E)}
    \State{\textbf{Input:} $G_{init}$; The beam search width $W$;}
    \State{\textbf{Output:} \revision{A set of optimal \textit{flat OPP graphs} $\mathbb{O}$;}}
    \revision{
    \Statex// Data structure setting
    \State{A directed acyclic graph $G_{solution}$ to encode the solution space;}
    \State{A nodes vector $\mathbb{P}$ of the previous depth of beam search;}
    \State{A nodes vector $\mathbb{C}$ of the current depth of beam search;}
    \Statex// Data structure initialization    
    \State{$\mathbb{O}$ $\gets$ $\varnothing$; $\mathbb{P}$ $\gets$ $\varnothing$; $\mathbb{C}$ $\gets$ $\varnothing$;}
    %an empty node (an additional virtual node)
    \State{Set $G_{solution}$'s root node to an empty virtual node;}
    \State{Add $G_{solution}$'s root node to $\mathbb{P}$;}
    \Statex// Solution space exploration with the beam search strategy  
    \While{$\mathbb{P} \ne \varnothing$}
        \Statex$\quad \ \ $// Generate candidate nodes of current depth from $\mathbb{P}$
        \State{Declare a set of node sequences $\mathbb{S}$; $\mathbb{S} \gets \varnothing$; $\mathbb{C}$ $\gets$ $\varnothing$;}
        \For{each node $\blacklozenge$ of $\mathbb{P}$}
            \State{Get\;the\;node\;sequence\;$s_i$\;from\;$G_{solution}$'s\;root\;node\;to\;$\blacklozenge$;}
            \State{Get the corresponding \textit{flat OPP graph} $G_{flat}$ of $s_i$;}
            \State{Get the node set $M_i$ of $G_{init}$ which are included in $s_i$;}
            \If{$M_i$ == all nodes of $G_{init}$ $\&$ $G_{flat} \notin \mathbb{O}$}
            \State{$\mathbb{O} \gets \mathbb{O} \cup \{G_{flat}\}$;}
            \State{Continue;}
            \EndIf
            %\Statex{$\quad\ \ $//Find the nodes in next iteration, and store them to $\mathbb{C}$}
            \Statex$\quad \quad \quad$// Generate candidate nodes from $\blacklozenge$
            %\State{Get the edge set $E_i$ of $G_{init}$ which are related to $M_i$;}
            \State{Get $\overline{G_{init}}$ from $G_{init}$ by deleting related edges to $M_i$;}
            %\State{Get the indegree vector $I_i$ of nodes in $\overline{G_{init}}$;}
                \For {each node $\overline{n_j} \in \overline{G_{init}}$}
                    \If{$!(\overline{n_j}'s$ indegree $ == 0$ and $\overline{n_j} \notin M_i)$}
                        \State{Continue;}
                    \EndIf
                    %\State{DFS from $\overline{n_j}$ in $\overline{G_{init}}$, get a set of paths;}
                    \State{Explore all traversal paths $\overline{P_j}$ starting from $\overline{n_j}$ with the greedy strategy to explore $\overline{G_{init}}$ as deep as possible;}
                    \Statex$\quad \quad \quad \quad \quad$// Update $G_{solution}$ and $\mathbb{C}$
                    \For{each traversal path $\overline{p_k} \in \overline{P_j}$}
                    \State{Take $\overline{p_k}$ as a new candidate node $\blacksquare$ of $G_{solution}$;}
                    \State{$b \gets$ whether a sequence exists in $\mathbb{S}$ with the same $G_{init}$'s nodes as $(s_i,\blacksquare)$;}
                    \State{Declare the last node of selected sequence $\blacktriangle$};
                    \If{b $\&$ $(\blacksquare == \blacktriangle)$}
                        \State{Add an edge from $\blacklozenge$ to $\blacktriangle$ in $G_{solution}$;}
                    \Else
                        % add edge add nodes add nodes to vc
                        \State{Add $\blacksquare$ and an edge from $\blacklozenge$ to $\blacksquare$ in $G_{solution}$;}
                        \State{$\mathbb{C}$ $\gets$ $\mathbb{C} \cup \{\blacksquare\}$;}
                    \EndIf
                    \State{$\mathbb{S}$ $\gets$ $\mathbb{S}$ $\cup$ $\{(s_i,\blacksquare)\}$;}
                    \EndFor
                \EndFor
        \EndFor
        \Statex$\quad \ \ $// Apply the branch and bound technique
        \If{$\mathbb{O} \ne \varnothing$}
        \State{Return $\mathbb{O}$;}
    	\EndIf
    	\Statex$\quad \ \ $// Extract $W$ nodes from candidate nodes of current depth
        \State{Sort $\mathbb{C}$ by the number of included $G_{init}$'s nodes;} 
        \State{$\mathbb{P}$ $\gets$ the first $W$ nodes of $\mathbb{C}$;}
    	%\State{Retain the nodes of $\mathbb{P}$ in $G_{solution}$, and delete other \Statex{$\quad\quad\quad$nodes in current depth;}}
    \EndWhile
    %\State {{\color{blue}\Return The remaining shortest paths, each path is a MPC.}}
    }
\end{algorithmic}
\end{algorithm}

%Note that nodes of $G_{solution}$ would point to the same node of next level if its sequence included $G_{init}$'s nodes remain the same, such as \{(1,2), (4)\} and \{(1,4), (2)\} both point to (3,6) and (3,5).}

%This is a process in which depth-first search (DFS) is nested in breadth-first search (BFS). 
%BFS builds the $solution{\_}space$ ($ss$). 
%DFS finds each node on the $ss$ (r13). 
%Before searching for next nodes from the current node in $ss$, we delete paths in $G_{init}$ that have been searched from the current node to the root of $ss$ (r8). After searching the current node is over, the opposite operation needs to be done to avoid interfering next search (r15).
%After searching all nodes is over in the current layer of $ss$, choose W new searched nodes to update $ss$ (r20-22). 
%When all nodes in $G_{init}$ are visited, we will not perform searching of the next layer (r18).\hs{what is r ? \yl{row}}

%\renewcommand{\algorithmicrequire}{\textbf{Input:}}  % Use Input in the format of Algorithm
%\renewcommand{\algorithmicensure}{\textbf{Output:}} % Use Output in the format of Algorithm

\begin{algorithm}[t]
    \caption{OPP Merging through Curved Layers}
    \label{alg:four}
    \begin{algorithmic}[1]
    %\REQUIRE {A directed acyclic graph G(V,E)}
    \State{\textbf{Input:} \revision{A $G_{flat}$, $G_{init}$;}}
    \State{\textbf{Output:} \revision{A $G_{curved}$;}}
    \revision{
    \State{Two nodes $n_l$, $n_r$ of $G_{curved}$;}
    \State{$G_{curved}$ $\gets$ $G_{flat}$;}
    \Statex{// Initial merging process}
    \For{each node $\blacklozenge$ of $G_{curved}$}
        \State{Pairwisely merge sub-OPPs of $\blacklozenge$ with \textit{curving} operation;}
    \EndFor
    \Statex{// OPP merging process}
    \State{$b \gets True$}
    \While{b}
    \State{$b \gets False$}
        \For{each edge ($n_l$, $n_r$) of $G_{curved}$}
            \Statex{$\quad \quad \ \ \ \ $// Deadlock detection}
            \If{more than one path between $n_l$ and $n_r$}
                \State{Continue;}
            \EndIf
            \Statex{$\quad \quad \ \ \ \ $// Inner Loop of OPP Merging Process}
            \State{Formulate a DAG $G_{sub}$ from sub-OPPs of $n_l$ and $n_r$:}
            \State{$\quad$ Get two sub-OPP sequences of $n_l$ and $n_r$;}
            \State{$\quad$ Add back the edges between two sequences in $G_{init}$;}
            \State{Pairwisely merge nodes of $G_{sub}$ with \textit{curving} operation;}
            \State{$\quad$ Label the nodes that have edges across the sequences;}
            \State{$\quad$ Select edges of $G_{sub}$ over such labeled nodes;}
            \State{$\quad$ For each selected edge, call \textit{curving} operation;}
            \Statex{$\quad \quad \ \ \ \ $// Update $G_{curved}$}
            \If{$G_{sub}$ has been merged to a single OPP node}
                \State{Update $G_{curved}$ by merging $n_l$ and $n_r$;}
                \State{Update sub-OPP sequences of the merged OPP node;}
                \State{$b \gets True$;}
                \State{Break;}
            \EndIf
        \EndFor
    \EndWhile}
\end{algorithmic}
\end{algorithm}

%\newpage{}
%\clearpage
%\input{LanguageSup.tex}

\end{document}